\documentclass[12pt ]{article}
\usepackage{jheppub_kim}
  \usepackage [ ]{inputenc}
 \usepackage{subcaption}
 
\usepackage{amsmath}
\usepackage{amssymb}
\usepackage{dcolumn}
\usepackage{bm}
\usepackage{color}
\usepackage{epsfig}
\usepackage{amsfonts}
\usepackage{graphicx} 
  
\begin{document}
 \title{Non-perturbative corrections to  quasi-topological black hole thermodynamics}
 \author[a]{Vivek Kumar Srivastava,}
 \author[b,c,d]{Sudhaker Upadhyay,\footnote{Corresponding author}\footnote{Visiting Associate, Inter-University Centre for Astronomy and Astrophysics (IUCAA) Pune-411007, Maharashtra, India}}
 \author[a]{Alok Kumar Verma,}
    \author[e,c]{Dharm Veer Singh,\footnote{Visiting Associate, Inter-University Centre for Astronomy and Astrophysics (IUCAA) Pune-411007, Maharashtra, India}} 
 
 \author[d]{Yerlan Myrzakulov,}
  \author[d]{Kairat Myrzakulov,}
        \affiliation[a]{Department of Physics, Prof. Rajendra Singh (Rajju Bhaiya) Institute of Physical Sciences for Study and Research,
Veer Bahadur Singh Purvanchal University, Jaunpur, 222001, Uttar Pradesh, India}
   \affiliation[b]{Department of Physics, K.L.S. College, Nawada, Magadh University, Bodh Gaya, Bihar 805110,  India}
    \affiliation[c]{School of Physics, Damghan University, P.O. Box 3671641167,  Damghan, Iran}
    \affiliation[d]{Department of General \& Theoretical Physics, L. N. Gumilyov Eurasian National University,  Astana, 010008, Kazakhstan}
       	 \affiliation[e]{Department of Physics, GLA University, Mathura, Uttar Pradesh 281406, India}
 
       \emailAdd{shubhoniam.1312@gmail.com} 
 \emailAdd{sudhakerupadhyay@gmail.com; sudhaker@associates.iucaa.in}
  \emailAdd{alok9369@gmail.com} 
   \emailAdd{veerdsingh@gmail.com} 
 \emailAdd{ymyrzakulov@gmail.com}
  \emailAdd{krmyrzakulov@gmail.com}

\abstract{We examine the impact of non-perturbative quantum corrections to the entropy of both charged and charged rotating quasi-topological black holes, with a focus on their thermodynamic properties. The negative-valued correction to the entropy for small black holes is found to be unphysical. Furthermore, we analyze the effect of these non-perturbative corrections on other thermodynamic quantities, including internal energy, Gibbs free energy, charge density, and mass density, for both types of black holes. Our findings indicate that the sign of the correction parameter plays a crucial role at small horizon radii. Additionally, we assess the stability and phase transitions of these black holes in the presence of non-perturbative corrections.
Below the critical point, both the corrected and uncorrected specific heat per unit volume are in an unstable regime. This instability leads to a first-order phase transition, wherein the specific heat transitions from negative to positive values as the system reaches a stable state.}
\keywords{thermodynamics; quantum corrections; horizon radius; specific heat.} 
\maketitle
\section{Introduction}\label{intro}
Gauge/gravity duality correlates the gravity theory with the gauge theory, which was first embraced by the AdS/CFT correspondence and proposed by Maldacena \cite{Maldacena1999}.
This correspondence has been used for making the enormous strides to understand quantum field theories and quantum gravity. It also imparts a novel framework to the holographic study in non-perturbative regimes. 
In the bulk space-time, this duality for the Einstein general relativity corresponds to a gauge theory for large $N_c$ (number of colors) and large $\lambda$ ('t Hooft coupling) \cite{Witten1998, Gubser1998}.
Nevertheless, within a perturbative frame to extend our consideration to the finite $N_c$ and finite $\lambda$ corrections, it is required to append the higher derivative terms or higher curvature terms with more coupling constants \cite{Buchel2005, Myers2010holographic}. 
The existence of different higher-order derivatives in AdS gravity correlates to the emergence of novel connections among operators in the dual conformal field theory (CFT).
Ab initio, abstraction is to add one of the higher derivative gravity theories as a Gauss-Bonnet term. 
Notwithstanding the inclusion of the Gauss-Bonnet term, dual theory remains limited due to its only one quadratic coupling term.
In order to improve the limitation, high order curvature cubed interaction term has been introduced into the Lovelock gravity \cite{Hofman2008, Hofman2009}. 
The third order Lovelock gravity also has constraints for four dimensional field theories because of its second order differential equation of motion. In the context of third order Lovelock gravity, disappearance of cubic component is brought about by the topological origin of Euler density in six dimensions. 
In sight of removing such discrepancy, a new toy model of curvature 
cubed interaction term along with the Gauss-Bonnet term was included 
with its contribution in five dimensions. For the spherically 
symmetric metric, this new model provides the second order 
differential equations of motion. This theory does not have a 
topological origin like Lovelock model, so it has been proposed as 
the quasi-topological gravity \cite{Myers2010black, Oliva2010}. 

The holographic and basic thermodynamical study of quasi-topological 
black hole have been performed in Ref \cite{Myers2010holographic, 
Brenna2012}. 
The introduction of surface term into quasi-topological gravity was 
intended to accentuate the well-definedness of the variational 
principle of action. Additionally, other thermodynamical parameters 
and the first law of thermodynamics  have been examined through the 
application of Gibbs free energy. Subsequently, solutions were 
generalized to encompass the rotating charged case 
\cite{Dehghani2011surface}. 
The solutions corresponding to Einstein equations of a black hole are the function of area   of the black hole and surface gravity  of the black hole, which imparts a close resemblance with black hole entropy and temperature, respectively \cite{Bardeen1973, Bekenstein1973, Hawking1976}. 
The holographic principle allows one to correlate the emitted radiation from black hole as a black body with its characteristic entropy related to the area of black hole, as the equilibrium entropy is quarter of the area of black hole numerically \cite{Mathur2012, Pourhassan2021a, Pourhassan2021b}.
The thermodynamic behavior of black holes is one of the most intriguing aspects, as it is described through analogies to classical thermodynamics. 
The Smarr formula is the relation of parameters like mass, charge, and angular momenta associated with the black hole, as the first law of black holes has been generalised for both static and rotating black holes \cite{Gulin2018}.
The classical and quantum scenarios of black holes provide different perspectives based on the microscopic origin. The introduction of Hawking radiation and information paradox yielded from the quantum aspects of black hole physics.
The radiation, termed Hawking radiation, emitted from black holes causes decrease in the size of black holes and becomes stable at quantum ground level \cite{Page2005, Ashtekar2020}.

The thermal and statistical fluctuations provide the microscopic origin of entropy, which may be interpreted as one of the most effective methods for analyzing the quantum-level correction \cite{Upadhyay2018, Dehghani2019, PourhassanUpadhyay2021}. The thermal fluctuation increases the disorderness in the system.
With the decrease in the size of black hole due to the Hawking radiation, leading-order perturbative quantum corrections in the form of second order corrections and logarithmic corrections come across \cite{Pourhassan2021}.
Recently, the perturbative quantum corrections of the charged quasi-topological and charge rotating quasi-topological black holes due to thermal fluctuations have been studied \cite{Upadhyay2017quantum}. 
However, due to the Hawking radiation, further decrease in the size of black hole to Planck scale, non-perturbative quantum corrections become dominant over the perturbative corrections.  
In the quantum theory of gravity, an exponential term proposed the non-perturbative corrections in the black hole entropy is pertinent to the other thermodynamical variables as well \cite{Chatterjee2020, Soroushfar2023, non2024pourhassan}.

In section \ref{ecbhe}, we discuss the  non-perturbative corrections due to the thermal fluctuations based on the quantum effects for a black hole entropy. 
In section \ref{cqtbh}, we study the higher derivative gravity theories generalised by Gauss-Bonnet and Lovelock gravity   in $(n+1)$ space-time dimensions.
Further on, we focus on the gravitational theory of $(n+1)$ space-time dimensions for a quasi-topological gravity in the presence of an electromagnetic field. The exact solutions corresponding to both charged and charged rotating quasi-topological black holes have been analysed on different thermodynamical parameters due to the thermal fluctuations.
Proceeding further on adhering to the standard methodologies, we discuss   the equilibrium entropy of charged quasi-topological black hole using Gibbs free energy and Hawking temperature in the absence of thermal fluctuations.
Within this section, we analyse the non-perturbatively corrected entropy and other thermodynamical parameters associated with the thermal fluctuations for charged quasi-topological black hole.
We discuss the stability and phase transition   for the charged   quasi-topological black holes also. Based on these findings, we also evaluated the bound points for both black holes.
We carry out the similar study of non-perturbative corrections for the charged rotating quasi-topological black hole in the presence of thermal fluctuations in section \ref{crqtbh}. 
In last section \ref{last}, we squeeze out the conclusion with the comparative analysis based on the effect of positive and negative correction parameters in non-perturbative analysis with the uncorrected parameters.

 \section{Exponential corrected black hole entropy}
\label{ecbhe}
The findings of the conducted research thus far infer that the horizons of black holes are same as those of thermodynamic systems, and the entropy of a black hole is commensurate with the horizon area. Classical black holes do not emit radiation, which means they are stable. Whereas quantum mechanically, the radiation emitted from the black hole, which may yield evaporation, decreases the size of black hole, called Hawking radiation \cite{Xu2020, Emparan2023}. This emission of radiation opens up the possibilities for final stage of the black hole, as either all of its mass will radiate or it will attain a specific stage at the point of stability. Accordingly, stable small black holes, compared to Planck scale, necessitate corrections with their entropy based on the perturbative and non-perturbative quantum effects \cite{Pourhassan2021, Pourhassan2022}. 
The modification of holographic principle results from the quantum effect in the form of a correction factor that emerges from thermal fluctuations \cite{Rama1999, Bak2000}.
In a thermodynamical system, leading order perturbative corrections to the black hole entropy are propounded by the logarithmic term \cite{Kaul2000, Dabholkar2011, Upadhyay2017} or power law for higher dimensions \cite{Iorio2020} and likely for higher order corrections \cite{Pourhassan2018}.
However, black hole entropy gets corrected with the exponential term by using the non-perturbative quantum theory of general relativity for the case when microstate counting is performed for the quantum states residing on the horizon only. 
The unmeasurable microstates, which contribute to the entropy, are inversely proportional to the number of measurable microstates \cite{Chatterjee2020, Soroushfar2024}, providing the entropy as 
\begin{align}
S_{micro} = e^{-S_0}.
\end{align}
This non-perturbative analysis would dominate in the case when more reductions in black hole size have been identified.
The total entropy provided for all the quantum corrections is given as
\begin{equation}
\label{eqtentropy}
S = S_0 + \alpha \ln S_0 + \frac{\lambda}{S_0} + \eta e^{-S_0},
\end{equation}
where $S_0$ is the original entropy, $\alpha$, $\lambda$, and $\eta$ are some infinitesimal constants, called as correction coefficients.
The second and third terms of Eq. (\ref{eqtentropy}) provides the perturbative correction, whereas the fourth term is associated with the non-perturbative quantum corrections \cite{Chatterjee2020, Soroushfar2023}.

This work is completely oriented towards the non-perturbative correction of the quasi-topological black hole. The non-perturbatively corrected black hole entropy is solely expressed with an exponential term as follows
\begin{align}
\nonumber
S &= S_0 + S_{micro}, \\\label{eqnpentropy}
&=S_0 + \eta e^{-S_0}.
\end{align}  
The general correction parameter $\eta$ of the above expression do not possess a fixed value, it will take the different values based on the different criteria and conditions. 
From Eq. (\ref{eqnpentropy}), one may conclude that for large horizon area $(A)$, exponential factor is negligible because of the fact that for $A\gg1$ we have $S_0\gg1$, hence $S=S_0$ \cite{Pourhassan2021, Pourhassan2022}. Another noteworthy observation is that, despite the disappearance of horizon area, overall entropy remains nonzero.
In this work, we will exclusively study the charged and rotating charged quasi-topological black holes under the influence of exponentially corrected entropy.  
\section{Charged quasi-topological black holes}
\label{cqtbh}
The extension of quasi-topological gravity to the higher dimensions, in (n+1) space-time dimensions, provides the general form of action in the following manner \cite{Myers2010black, Bueno2020generalised}
\begin{align}
\label{actiongen}
I = \frac{1}{16 \pi G_{n+1}} \int \,d^{n+1} x \sqrt{-g} \left[ R - 2\Lambda + \lambda l^2 \mathcal{Z}_2 + \mu l^4\mathcal{Z}_3 \right],
\end{align}
where first term $R$ is well known Einstein-Hilbert
action and second term $\Lambda$ is the cosmological constant defined as $\Lambda = -\frac{n(n-1)}{2l^2}$ where $l$ is the characteristics length scale of the AdS space, referred to as AdS radius. $\lambda$ and $\mu$ are the Gauss-Bonnet and quasi-topological coupling constant, respectively. The term $\mathcal{Z}_2$ is Gauss-Bonnet term and $\mathcal{Z}_3$ is quasi-topological term, be expanded as
\begin{align}
\mathcal{Z}_2 &= \frac{1}{(n-2)(n-3)}\chi_{_2}, \\
\mathcal{Z}_3 &= \frac{8 (2n-1)}{(n-2)(n-5)(3n^2-9n+4)} \chi_{_3}.
\end{align}
The term $\chi_{_2}$ corresponds to the Gauss-Bonnet theory of gravity for the Euler characteristic of 4-D manifolds which associates the curvature-squared interaction and form of density as
\begin{align}
\chi_{_2} = R^2  - 4 R_{\mu\nu} R^{\mu\nu} + R_{\mu\nu\eta\sigma} R^{\mu\nu\eta\sigma},
\end{align}
and $\chi_{_3}$ of the quasi-topological term is in the form of
\begin{align}
\nonumber
\chi_{_3} &= R_{\alpha~\beta}^{~\eta\sigma} R_{\eta\sigma}^{~\gamma\delta} R_{\gamma\delta}^{~\alpha~\beta} + \frac{1}{(2n-1)(n-3)} \left[ \frac{3(3n-5)}{8}R_{\mu\nu\eta\sigma} R^{\mu\nu\eta\sigma}R - 3(n-1) R_{\mu\nu\eta\sigma} R^{\mu\nu\eta}_{~~~~\alpha}R^{\sigma\alpha} \right.\\&\left. + 3(n+1) R_{\mu\nu\eta\sigma} R^{\mu \eta} R^{ \nu\sigma} +6(n-1) R_{\mu}^{~\nu} R_{\nu}^{~\eta} R_{\eta}^{~\mu} - \frac{3(3n-1)}{2} R_{\mu}^{~\nu} R_{\nu}^{~\mu} R + \frac{3(n+1)}{8}R^3  \right].
\end{align}
The Lagrangian of third-order Lovelock gravity does not contribute to the field equation in five dimensions, whereas the quasi-topological term contributes in five dimensions. However, quasi-topological terms do not contribute to the field equations in six or less than five-dimensional space-time \cite{Lovelock1971the, Deruelle1990lovelock}. In the static metric case, quasi-topological gravity corresponds to the second-order differential field equation \cite{Dehghani2011surface}. 
 
Additionally, surface parameters contribute to the well-defined variation in the action of quasi-topological gravity. 
The general action, Eq. (\ref{actiongen}), modifies in the presence of an electromagnetic field in $(n+1)$ dimensions for quasi-topological gravity as \cite{Dehghani2011surface, Upadhyay2017quantum}
\begin{align}
\label{actionem}
I = \frac{1}{16 \pi G_{n+1}} \int \,d^{n+1} x \sqrt{-g} \left[ R - 2\Lambda + \lambda l^2 \mathcal{Z}_2 + \mu l^4\mathcal{Z}_3 - \frac{1}{4}F_{\mu\nu}F^{\mu\nu} \right],
\end{align}
where $F_{\mu\nu} = \partial_\mu A_\nu - \partial_\nu A_\mu$ is Maxwell field-strength tensor and $A_\nu$ is vector potential. It is well known that in the case of static solutions, quasi-topological gravity endows the second-order differential equations.
In this study, we adequately define the action (\ref{actionem}) for the static metric by introducing the surface term, using the similarity between quasi-topological gravity and third-order Lovelock gravity. To facilitate this process, one considers the case of space-time with a flat boundary  \cite{Dehghani2011surface}. Moreover, it is widely recognized that one can obtain a merely second-order differential equation of motion for quasi-topological gravity in the case of static and spherically symmetric (SSS) metric solutions \cite{Bueno2024regular}. In the perpetuity of this work, the SSS metric for $(n+1)$ dimensional  spacetime with a flat boundary \cite{Upadhyay2017quantum, Bueno2024regular, Filippo2024inner} is given as follows
\begin{align}
\label{metric}
\,d s^2 = -g(r) \,d t^2 + \frac{\,d r^2}{f(r)} + r^2 \sum_{i=1} ^ {n-1} \, d \phi_i^2,
\end{align}
where $g(r)=N^2(r) f(r)$ and $\displaystyle\sum_{i=1} ^ {n-1} \, d \phi_i^2$ represents the angular coordinates in $(n-1)$ dimensional spherical geometry. 
The function $N(r)$ is a lapse function that must be constant and set to be unit $(\it i.e., N(r)=1)$ and varying the action with respect to h(r) provides the solution for metric function as \cite{Dehghani2011surface} 
\begin{align}
f(r) = \frac{r^2}{l^2} - \frac{m}{r^{d-2}} + \frac{q^2}{r^{2(d-2)}},
\end{align}
where $m$ and $q$ are the integration constant represents the total mass and electric charge of the quasi-topological black hole. The equation of motion provides the solution for $h(r)$ in the form of
\begin{align}
h(r) = -\sqrt{\frac{2(n-1)}{n-2}}\frac{q}{r^{n-2}}.
\end{align}
On the horizon $(f(r=r_+)=0)$, using the metric function one may calculate the integration constant $m$ as 
\begin{align}
m = \frac{r_+^n}{l^2}+\frac{q^2}{r_+^{n-2}},
\end{align}
where, $r_+$ is outer horizon radius.
Proceeding further on following the standard method of analytic continuation of the metric, one may obtain the temperature of the event horizon as follows
\begin{align}
\label{eqhtp}
T = {\left.\frac{f'(r)}{4 \pi}\right|_{r=r_+}} = \frac{nr_+ - (n-2)q^2l^2r^{3-2n}}{4\pi l^2}.
\end{align}
For the static case, electric potential at infinity with respect to the outer horizon can be expressed as follows \cite{Cvetic1999, Caldarelli2000, Dehghani2011surface, Upadhyay2017quantum}
\begin{align}
\label{eqpot}
\Phi=\sqrt{\frac{2(n-1)}{n-2}} \frac{q}{r_+^{n-2}}.
\end{align}
Now, using the Eqs. (\ref{eqhtp}) \& (\ref{eqpot}), the Hawking temperature of the quasi-topological  black hole be expressed as \cite{Upadhyay2017quantum}
\begin{align}
\label{eqhtemp}
T_H &= \frac{nr_+}{4\pi l^2} - \frac{\left(n-2\right)^2 \Phi^2}{8\pi\left(n-1\right)r_+},
\end{align}
and the horizon radius can be expressed in terms of  temperature $(T)$ and electric potential $(\Phi)$ as follows 
\begin{align}
\label{eqradius}
r_+&= \frac{2\pi l^2 T}{n} + 2 \left[ \frac{\pi^2 l^4 T^2}{n^2} + \frac{\left(n-2\right)^2 l^2 \Phi^2}{8n\left(n-1\right)} \right]^{1/2}.
\end{align}
Further on, Gibbs free energy per unit volume, which results from the Euclidean action per volume times temperature, leads to the entropy density of charged quasi-topological black hole as in the form of
\begin{equation}
\label{eqentropy}
S_0 = \frac{r^{n-1}_+}{4}.
\end{equation}
Due to the thermal fluctuations, non-perturbatively corrected entropy per volume for the charged quasi-topological black hole can be ascertained by using the Eqs. (\ref{eqnpentropy}) and (\ref{eqentropy}) as
\begin{equation}
\label{eqcentropy}
 S = \frac{r^{n-1}_+}{4} + \eta e^{-\frac{r^{n-1}_+}{4}}.
\end{equation}
\begin{figure}[ht]
     \centering
     \begin{subfigure}[b]{0.45\textwidth}
         \centering
         \includegraphics[width=\textwidth]{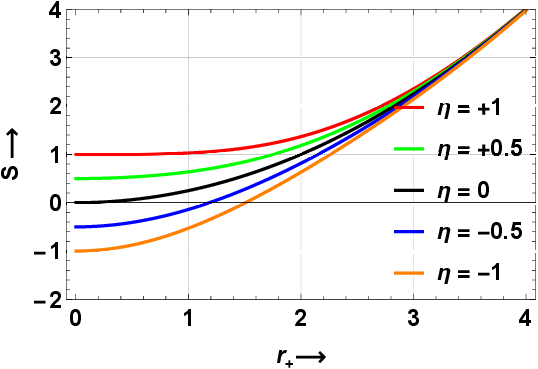}
         \caption{For $n = 3$}
         \label{entropy3}
     \end{subfigure}
     \hfill
     \begin{subfigure}[b]{0.45\textwidth}
         \centering
         \includegraphics[width=\textwidth]{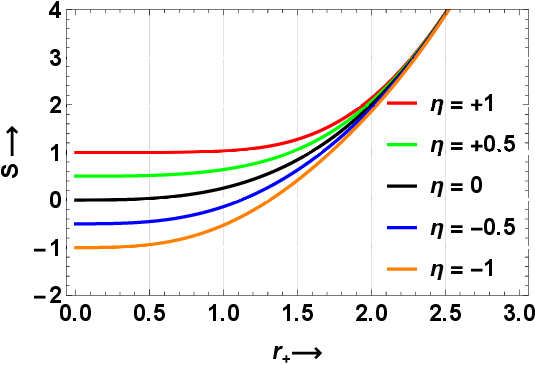}
         \caption{For $n = 4$}
         \label{entropy4}
     \end{subfigure}
           \caption{Variation of entropy per volume $(S)$ with respect to black hole horizon radius $(r_+)$   at different values of correction parameter $(\eta)$.}
\label{entropy} 
\end{figure}
 In Fig. \ref{entropy}, behavior of entropy per volume with respect to horizon radius is illustrated by the effects of non-perturbative correction. This variation of entropy density behaves as a monotonically increasing function.
The variations of entropy per volume have negligible effect in its nature, with the change of its dimension $(n)$. 
The positive value of correction parameter $(\eta)$ seized up with the positive value of entropy per volume for all ranges of the horizon radius in the similar manner as followed by the original equilibrium entropy.
Additionally, for the negative value of correction parameter, entropy density possesses negative value at lower horizon radius. However, as the horizon radius increases, the negative value of entropy density shifts towards the positive value for negative value of the correction parameter. For larger value of horizon radius, correction parameter becomes insignificant and converges to a particular asymptotic point. This converging asymptotic point shifted to the lower value of the horizon radius with an increase in the dimension $(n)$. 
 \subsection{Internal energy}
In the presence of non-perturbatively corrected entropy $(S)$  Eq. (\ref{eqcentropy})  for quasi-topological black hole, first-law of thermodynamics is in the form of 
\begin{align}
\,d E &=T_H \,d S,
\end{align}
which eventually leads to the internal energy of this system as
\begin{align}
E &=\int T_H \,d S.
\end{align}
Henceforth, using the Eqs. (\ref{eqhtemp}) \& (\ref{eqcentropy}), internal energy of the charged quasi-topological black hole is
\begin{align}
\label{eqinenergy}\nonumber
E &=\frac{1}{16 \pi l^2}\left[\left(n-1\right)r^n_+ - \frac{1}{2} \left(n-2\right) \phi^2 l^2 r_+^{n-2} + \eta \left\{ n4^{^{\frac{n}{n-1}}} \Gamma\left[\frac{n}{n-1}, \frac{r_+^{n-1}}{4}\right]  \right.  \right. \\& - \left. \left. \frac{\left(n-2 \right)^2}{2\left(n-1 \right)} \phi^2 l^2 4^{^{\frac{n-2}{n-1}}} \Gamma\left[\frac{n-2}{n-1}, \frac{r_+^{n-1}}{4}\right] \right\} \right].
\end{align}
\begin{figure}[ht]
     \centering
     \begin{subfigure}[b]{0.45\textwidth}
         \centering
         \includegraphics[width=\textwidth]{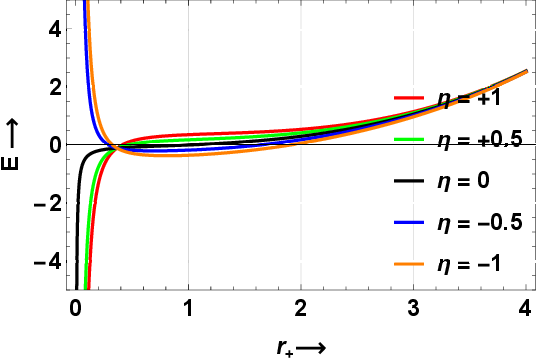}
         \caption{For $n = 3, q = 1, l = 1$}
    \label{energy311}
     \end{subfigure}
     \hfill
     \begin{subfigure}[b]{0.45\textwidth}
         \centering
         \includegraphics[width=\textwidth]{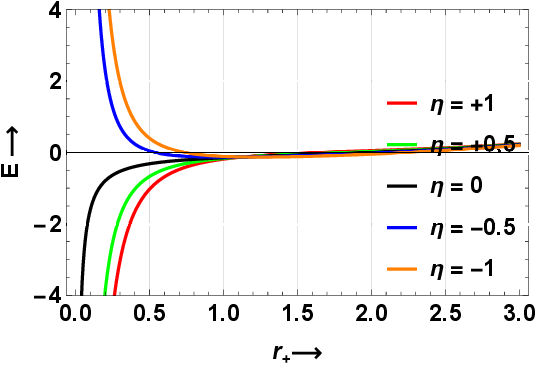}
         \caption{For $n = 3, q = 2, l = 2$ }
         \label{energy322}
     \end{subfigure}\\
        \begin{subfigure}[b]{0.45\textwidth}
         \centering
         \includegraphics[width=\textwidth]{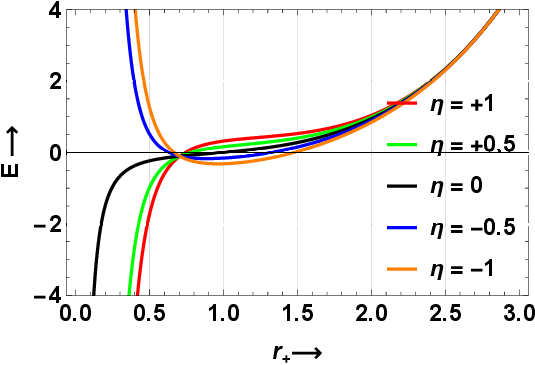}
         \caption{For $n = 4, q = 1, l = 1$}
       \label{energy411}
     \end{subfigure}
     \hfill
     \begin{subfigure}[b]{0.45\textwidth}
         \centering
         \includegraphics[width=\textwidth]{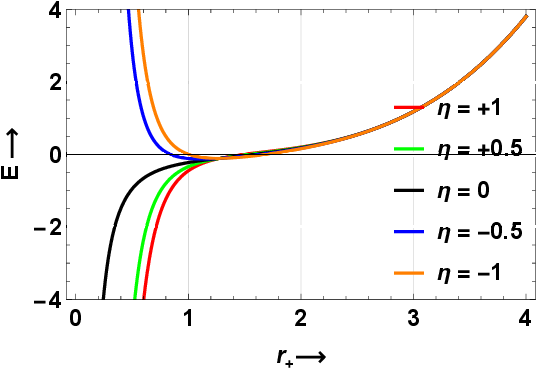}
         \caption{For $n = 4, q = 2, l = 2$}
        \label{energy422}
     \end{subfigure}
           \caption{Variation of internal energy $(E)$ with respect to black hole horizon radius $(r_+)$   at different values of correction parameter $(\eta)$.}
\label{energy} 
\end{figure}
 The comparable study of non-perturbatively corrected internal energy of the quasi-topological black hole with the equilibrium value has been studied from the plot shown in Fig. \ref{energy}. 
This variation with respect to the horizon radius $(r_+)$ elucidates a critical point at a point, say $r_+=r$. 
For the lower value of horizon radius, $(r_+<r)$, positive valued correction parameter possesses negative asymptotic behavior of internal energy, whereas the negative valued correction parameter holds the positive asymptotic behavior of internal energy. 
For the horizon radius $r_+>r$, only for the lower horizon radius the negative correction parameter bear the negative value. 
For the rest, larger horizon radius of negative correction parameter and all ranges of positive correction parameter for $r_+>r$, internal energy maintains the positive valued correction parameter. 
It is worth noting that the positive correction parameter shows similar behavior as that of the equilibrium value.
In the region of higher valued horizon radius, correction parameter $(\eta)$ does not play any significant role.
With an increase in the dimension $(n)$, nature of the plot does not change, while it becomes steeper and critical point shifts to the large value of internal energy. Whereas, with an increase in the charge and AdS radius, the variation effect because of the correction parameters shifted to the smaller horizon radius and graphs becomes less steep. 
\subsection{Gibbs free energy}
The other substantial thermodynamical quantity, Gibbs free energy per unit volume for charged quasi-topological black hole, can be evaluated by
\begin{equation}
\label{}
G(T_H, \Phi) =-\int S \,d T_H.
\end{equation}
Hence, using the Eqs. (\ref{eqcentropy}) \& (\ref{eqhtemp}) in above relation, Gibbs free energy per volume is as follows
\begin{align}
\label{eqgibbs} \nonumber
G &= - \frac{1}{16\pi l^2} \left[ r_+^n + \frac{(n-2)}{2(n-1)} \phi^2 l^2 r_+^{n-2} - \eta\left\{ \frac{n}{n-1} 4^{{\frac{n}{n-1}}} \Gamma\left[\frac{1}{n-1}, \frac{r_+^{n-1}}{4}\right]\right. \right. \\&\left. \left. + \frac{(n-2)^2}{2(n-1)^2} \phi^2 l^2 4^{{\frac{n-2}{n-1}}} \Gamma\left[\frac{-1}{n-1}, \frac{r_+^{n-1}}{4}\right] \right\} \right], \\
&= - \frac{1}{16\pi} \left[ m - \eta\left\{ \frac{n 4^{{\frac{n}{n-1}}}}{(n-1) l^2}  \Gamma\left[\frac{1}{n-1}, \frac{r_+^{n-1}}{4}\right] 
+ \frac{(n-2) q^2}{(n-1)r_+^{2n-4}} 4^{{\frac{n-2}{n-1}}} \Gamma\left[\frac{-1}{n-1}, \frac{r_+^{n-1}}{4}\right] \right\} \right].
\end{align}
\begin{figure}[ht]
     \centering
     \begin{subfigure}[b]{0.45\textwidth}
         \centering
         \includegraphics[width=\textwidth]{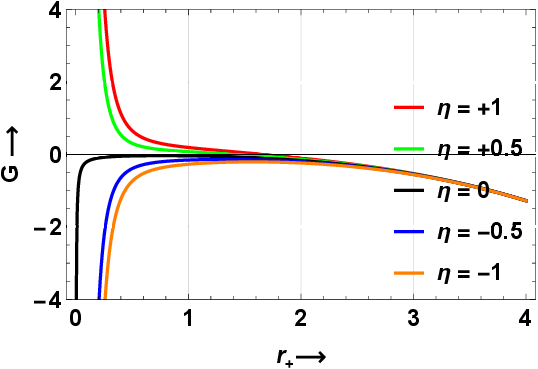}
         \caption{For $n = 3, q = 1, l = 1$}
    \label{gibbs311}
     \end{subfigure}
     \hfill
     \begin{subfigure}[b]{0.45\textwidth}
         \centering
         \includegraphics[width=\textwidth]{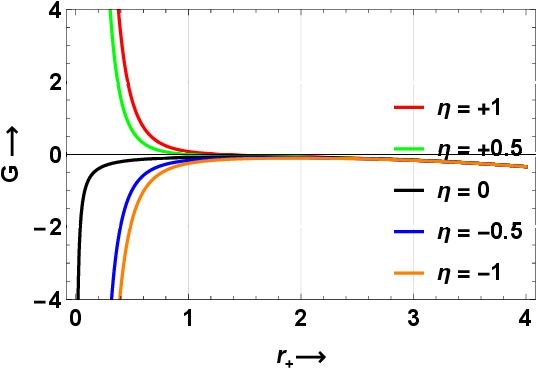}
         \caption{For $n = 3, q = 2, l = 2$ }
         \label{gibbs322}
     \end{subfigure}\\
      \begin{subfigure}[b]{0.45\textwidth}
         \centering
         \includegraphics[width=\textwidth]{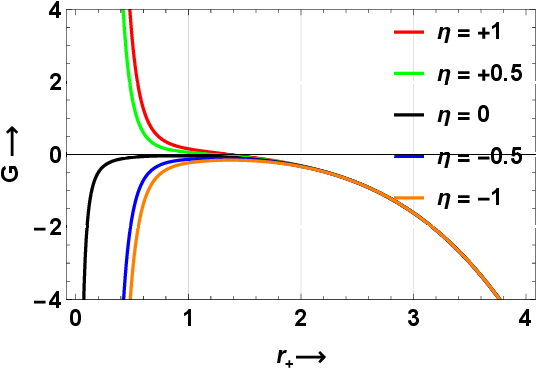}
         \caption{For $n = 4, q = 1, l = 1$}
       \label{gibbs411}
     \end{subfigure}
     \hfill
     \begin{subfigure}[b]{0.45\textwidth}
         \centering
         \includegraphics[width=\textwidth]{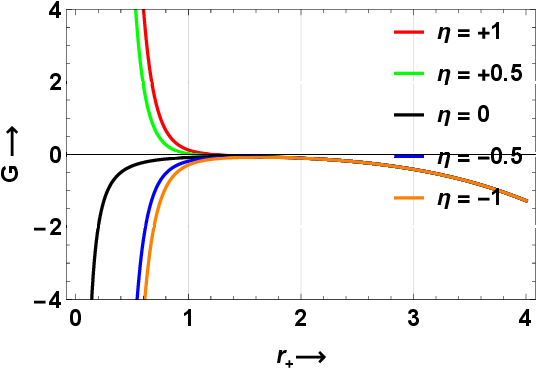}
         \caption{For $n = 4, q = 2, l = 2$}
        \label{gibbs422}
     \end{subfigure}
           \caption{Variation of Gibbs free energy per volume $(G)$ with respect to black hole horizon radius $(r_+)$ for  different values of correction parameter $(\eta)$.}
\label{gibbs} 
\end{figure}
 It is evident that for a limit $\eta \rightarrow 0$, Eq. (\ref{eqgibbs}) can be reduced in the original equilibrium form of Gibbs free energy per volume as obtained in Ref. \cite{Dehghani2011surface}. The comparative analysis of exponentially corrected and uncorrected Gibbs free energy per volume with respect to the horizon radius in four and five space-time dimensions can be seen in Fig. \ref{gibbs}. The positive and negative correction parameters possess positive and negative asymptotic values of Gibbs free energy per volume for charged quasi-topological black hole systems, respectively. The negatively corrected Gibbs free energy for the whole range of horizon radius follows the similar nature as followed by the uncorrected one. 
After the convergence point, this variation shows the negative valued Gibbs free energy and a decreasing function with an increase in horizon radius. It can also be seen that the correction parameter $(\eta)$ is only significant for the smaller value of horizon radius, while becoming insignificant for the large value of horizon radius. With an increase in the dimension $(n)$ and decrease in the charge $(q)$ and AdS radius $(l)$, variation develops into steeper form with the increase of horizon radius $(r_+)$.
\subsection{Charge density}
The exponentially corrected charged density of the charged quasi-topological black hole due to the thermal fluctuations can be interpreted as  
\begin{eqnarray}
\label{eqcharge} 
Q  = -\left(\frac{\partial G }{\partial \Phi}\right)_T 
  = \frac{1}{16 \pi} \sqrt{\frac{2(n-2)}{(n-1)}}q \left\{ 1- \eta  \frac{(n-2)4^{{\frac{n-2}{n-1}}}}{(n-1)r^{n-2}} \Gamma\left[\frac{-1}{n-1}, \frac{r_+^{n-1}}{4}\right] \right\}.
\end{eqnarray}
\begin{figure}[ht]
     \centering
     \begin{subfigure}[b]{0.45\textwidth}
         \centering
         \includegraphics[width=\textwidth]{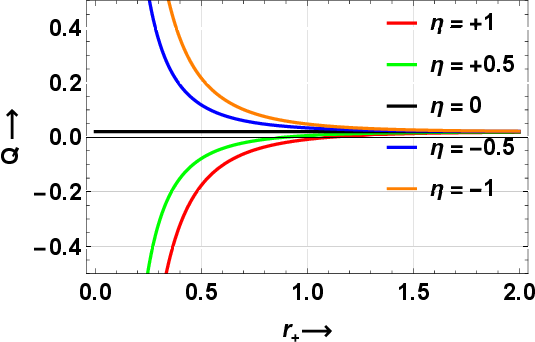}
         \caption{For $n = 3, q = 1, l = 1$}
    \label{charge31}
     \end{subfigure}
     \hfill
     \begin{subfigure}[b]{0.45\textwidth}
         \centering
         \includegraphics[width=\textwidth]{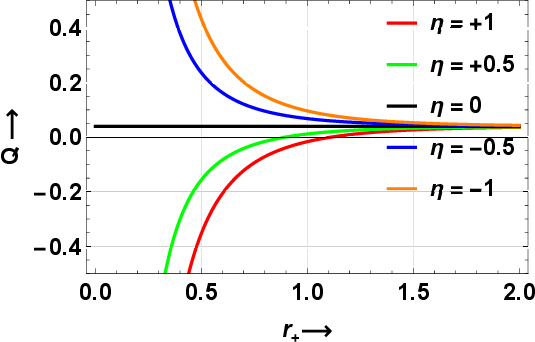}
         \caption{For $n = 3, q = 2, l = 2$ }
         \label{charge32}
     \end{subfigure}\\
      \begin{subfigure}[b]{0.45\textwidth}
         \centering
         \includegraphics[width=\textwidth]{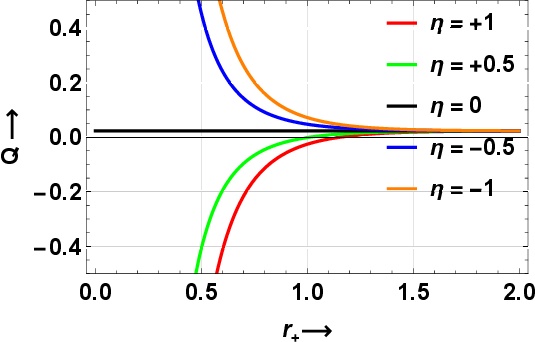}
         \caption{For $n = 4, q = 1, l = 1$}
       \label{charge41}
     \end{subfigure}
     \hfill
     \begin{subfigure}[b]{0.45\textwidth}
         \centering
         \includegraphics[width=\textwidth]{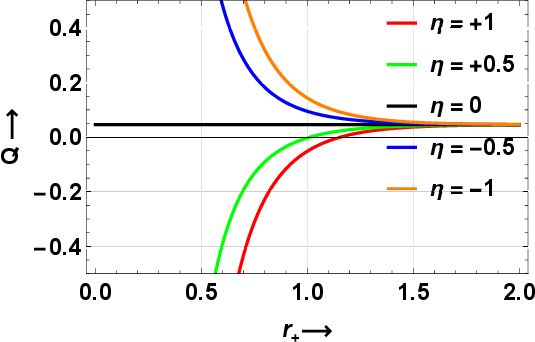}
         \caption{For $n = 4, q = 2, l = 2$}
        \label{charge42}
     \end{subfigure}
           \caption{Variation of charge $(Q)$ with respect to black hole horizon radius $(r_+)$ for different values of correction parameter $(\eta)$.}
\label{Charge} 
\end{figure} 
This exponentially corrected charged density can be obtained as calculated in Ref. \cite{Dehghani2011surface}, for a limiting case of $\eta \rightarrow 0$.
Notably, for less than four space-time dimensions, charge of the charged quasi-topological black hole does not exist and would also be futile for the exponential correction. 
The graphical variations of charge with respect to the horizon radius for four and five space-time dimensions are shown in Fig. \ref{Charge}. One may observe that the uncorrected charge is independent of the horizon radius. 
For negative and positive correction parameters, the charge variation is of positive and negative asymptotic behavior, respectively, for lower horizon radius. With the increase in horizon radius, both positive and negative exponentially corrected charged values decrease and seized to the uncorrected value and become invariant for large horizon radius. The change in dimension $(n)$, charge $(q)$, and AdS radius $(l)$ does not affect the nature of charge variation.
\subsection{Mass density}
\begin{figure}[ht]
     \centering
     \begin{subfigure}[b]{0.45\textwidth}
         \centering
         \includegraphics[width=\textwidth]{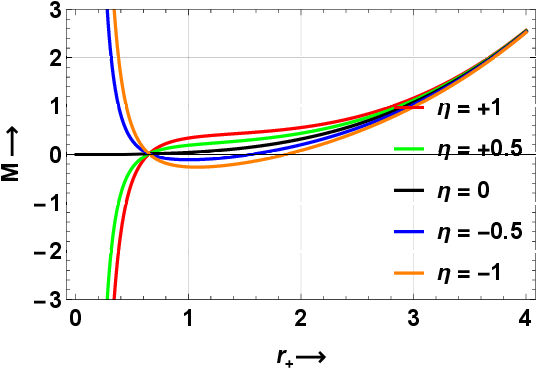}
         \caption{For $n = 3, q = 1, l = 1$}
    \label{mass311}
     \end{subfigure}
     \hfill
     \begin{subfigure}[b]{0.45\textwidth}
         \centering
         \includegraphics[width=\textwidth]{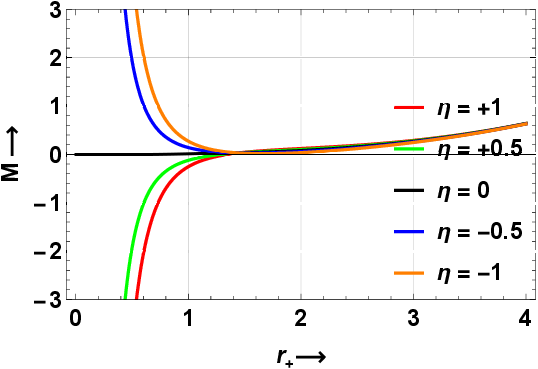}
         \caption{For $n = 3, q = 2, l = 2$ }
         \label{mass322}
     \end{subfigure}\\
      \begin{subfigure}[b]{0.45\textwidth}
         \centering
         \includegraphics[width=\textwidth]{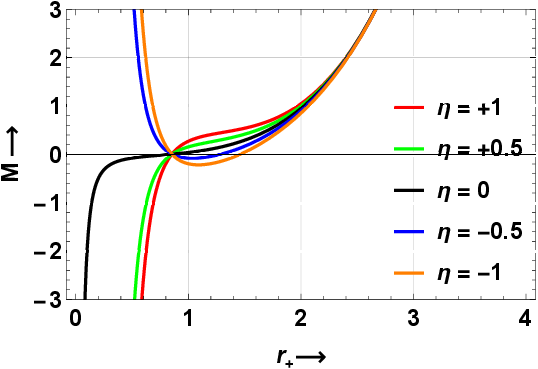}
         \caption{For $n = 4, q = 1, l = 1$}
       \label{mass411}
     \end{subfigure}
     \hfill
     \begin{subfigure}[b]{0.45\textwidth}
         \centering
         \includegraphics[width=\textwidth]{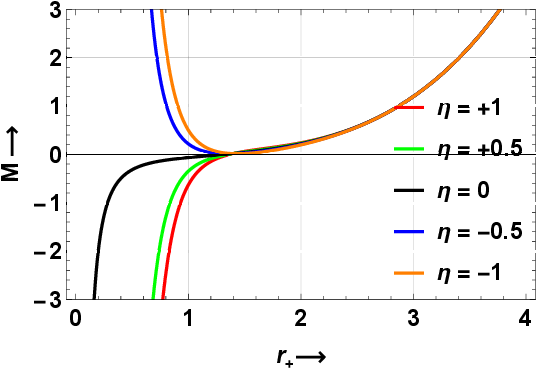}
         \caption{For $n = 4, q = 2, l = 2$}
        \label{mass422}
     \end{subfigure}
           \caption{Variation of mass per volume $(M)$ with respect to black hole horizon radius $(r_+)$ for  different values of correction parameter $(\eta)$.}
\label{mass} 
\end{figure}

Now, based on the different exponentially corrected parameters due to the thermal fluctuations, corrected mass density of the charged quasi-topological black hole can be calculated as
\begin{align}
\label{eqmass1} \nonumber
M &= G + TS + \Phi Q \\ \nonumber
& = \frac{1}{16 \pi l^2} \left[ (n-1)r_+^n +\frac{q^2 l^2 (3-n)}{r_+^{n-2}} + \eta \left\{ \frac{n}{n-1} 4^{^{\frac{n}{n-1}}} \Gamma\left[\frac{1}{n-1}, \frac{r_+^{n-1}}{4}\right] \right. \right. 
\\& \left. \left. -
\frac{(n-2)q^2 l^2}{(n-1)r_+^{2n-4}}
4^{^{\frac{n-2}{n-1}}} \Gamma\left[\frac{-1}{n-1}, \frac{r_+^{n-1}}{4}\right] + 4l^2 e^{-\frac{r_+^{n-1}}{4}} \left( \frac{nr_+}{l^2} - \frac{(n-2)q^2}{r_+^{2n-3}} \right)
\right\} \right].
\end{align}
The corrected mass density in term of mass of quasi-topological black hole be expressed as
\begin{align}
\label{eqmass2} \nonumber
M & = \frac{1}{16 \pi l^2} \left[ l^2 \left\{(n-1)m  +\frac{q^2 (4-2n)}{r_+^{n-2}} \right\}+ \eta \left\{ \frac{n}{n-1} 4^{^{\frac{n}{n-1}}} \Gamma\left[\frac{1}{n-1}, \frac{r_+^{n-1}}{4}\right] \right. \right. 
\\& \left. \left. -
\frac{(n-2)q^2 l^2}{(n-1)r_+^{2n-4}}
4^{^{\frac{n-2}{n-1}}} \Gamma\left[\frac{-1}{n-1}, \frac{r_+^{n-1}}{4}\right] + 4l^2 e^{-\frac{r_+^{n-1}}{4}} \left( \frac{nr_+}{l^2} - \frac{(n-2)q^2}{r_+^{2n-3}} \right)
\right\} \right].
\end{align}
The variation of both exponentially corrected and uncorrected mass per volume for the charged quasi-topological black hole can be comparably analysed from Fig. \ref{mass}. One may observe that this fluctuation shows a critical point at $r_+=r$ for small horizon radius. In five space-time dimensional space, positively corrected mass per volume follows the similar nature of trajectories as uncorrected mass per volume.  
However, in a four space-time dimensional space, nature of a positively corrected parameter persists with the variation as uncorrected mass density only beyond a critical point.
In the region of $0<r_+<r$, asymptotic behavior of corrected mass density with positive $\eta$ is completely opposite to that of corrected mass density with negative $\eta$. For $r_+>r$, for small size of black hole, negatively corrected mass per volume bears negative value. Whereas for sufficiently large horizon radius, mass density associated with both positive and negative correction parameters converges as a positively monotonically increasing function.
\subsection{Specific heat and Stability}
Thermal stability and phase during the Hawking evaporation process of the charged quasi-topological black holes can be analysed with the signature of specific heat. The divergence from a root value of specific heat will determine the phase transition of a particular black hole. This point is used to articulate the stable and unstable states. The negative value of specific heat represents an unstable thermodynamical system. From that juncture, an unstable state confronts a phase transition to the positive value of specific heat, which corresponds to a thermodynamically stable system.
\begin{figure}[ht]
     \centering
     \begin{subfigure}[b]{0.45\textwidth}
         \centering
         \includegraphics[width=\textwidth]{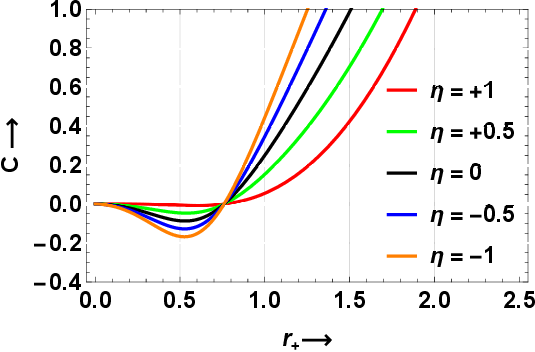}
         \caption{For $n = 3, q = 1, l = 1$}
    \label{sh311}
     \end{subfigure}
     \hfill
     \begin{subfigure}[b]{0.45\textwidth}
         \centering
         \includegraphics[width=\textwidth]{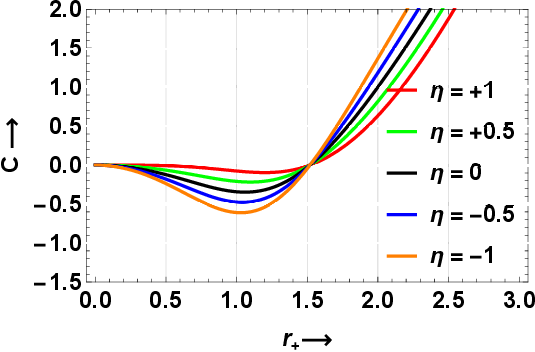}
         \caption{For $n = 3, q = 2, l = 2$ }
         \label{sh322}
     \end{subfigure}\\
      \begin{subfigure}[b]{0.45\textwidth}
         \centering
         \includegraphics[width=\textwidth]{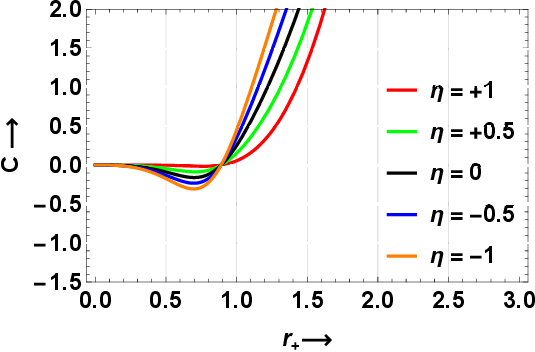}
         \caption{For $n = 4, q = 1, l = 1$}
       \label{sh411}
     \end{subfigure}
     \hfill
     \begin{subfigure}[b]{0.45\textwidth}
         \centering
         \includegraphics[width=\textwidth]{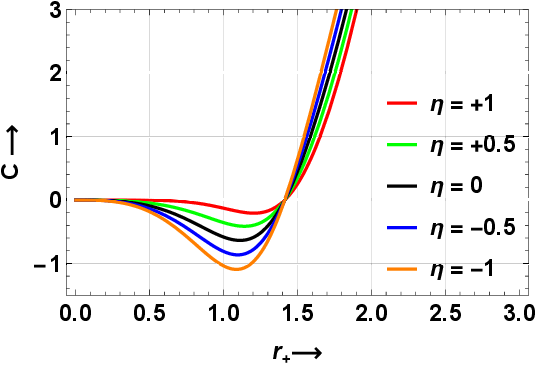}
         \caption{For $n = 4, q = 2, l = 2$}
        \label{sh422}
     \end{subfigure}
           \caption{Variation of Specific heat per volume $(C)$ with respect to black hole horizon radius $(r_+)$ for   different values of correction parameter $(\eta)$.}
\label{sh} 
\end{figure} 
Specific heat per volume of the thermodynamic system with a fixed chemical potential $(\Phi)$ is calculated by the relation as follows
\begin{align}
\nonumber
C_\Phi &= \left(\frac{\partial E }{\partial T}\right)_\Phi = T\left( \frac{\partial S }{\partial T}\right)_\Phi,
\\\nonumber
& = \frac{2\pi l^2 r_+^n (n-1)^2 T}{2n(n-1)r_+^2 +(n-2)^2 \phi^2 l^2}\left(1-\eta e^{-\frac{r_+^{n-1}}{4}}\right),\\
& = \frac{\left(n-1\right)r_+^{n-1} \left(\frac{2n}{l^2}- \frac{(n-2)^2\phi^2}{(n-1)r_+^2}\right)}{4 \left(\frac{2n}{l^2}+\frac{(n-2)^2\phi^2}{(n-1)r_+^2}\right)}\left(1-\eta e^{-\frac{r_+^{n-1}}{4}}\right).\label{eqsh}
\end{align}
The singularity calculated from the specific heat defines the bound points with respect to the horizon radius as
\begin{equation}
\label{eqstability}
r_{c} = \left[-\frac{n-2}{n}q^2 l^2 \right]^{1/2n-2}.
 \end{equation}
In Fig. \ref{sh}, there exists a critical point at $r_+=r$ for both the uncorrected as well as for positively and negatively corrected specific heat per volume, which can be calculated on solving the numerator of specific heat in respect of the horizon radius. 
The fluctuations of corrected and uncorrected specific heat exhibit negative values, leading to instability for a small horizon radius of quasi-topological black hole in the region of $0<r_+<r$. The critical point specifies the phase transition from unstable to stable state with positive value of specific heat per volume. In fact, for a sufficiently larger value of horizon radius for quasi-topological black hole, the corrected and uncorrected specific heat density is of positively asymptotic nature hence behaves as a stable system. 
Interestingly, correction parameter $(\eta)$, dimension $(n)$, charge $(q)$, and AdS radius $(l)$ does not play a significant role in phase transition, stability, or nature of the variation of specific heat per volume with respect to horizon radius.
\section{Charged rotating quasi-topological  black holes}
\label{crqtbh}
In this section, we will render the charged static solution with a global rotation to explicate the charged rotating quasi-topological black holes.
In $(n+1)$ dimensional asymptotically AdS rotating solution metric is defined as \cite{Dehghani2011surface, Upadhyay2017quantum}
\begin{align}
\nonumber
\,d s^2 &= -g(r) \left( \rho\,d t - \sum_{i=1}^k (a_i \,d\phi_i) \right)^2 + \frac{r^2}{l^4} \sum_{i=1}^k \left(a_i \,d t - \rho l^2 \,d \phi_i\right)^2 + \frac{\,dr^2}{f(r)} \\
& - \frac{r^2}{l^2} \sum_{i<j}^k \left( a_i \,d \phi_j -a_j \,d \phi_i \right)^2 + r^2 \sum_{i=k+1}^{n-1} \,d\phi_i^2,
\end{align}
for rotation parameters $k\leq \frac{n+1}{2}$, whose constant $(t,r)$ hypersurface has zero curvature \cite{Dehghani2006, Awad2019}. The angular coordinates range as $0\leq \phi_i\leq2\pi$ and $\rho$ is defined as
\begin{align}
\rho^2 = 1+ \sum_{i=1}^k\left(\frac{a_i}{l}\right)^2.
\end{align}
The factor $\rho^2$ can also be simplified as $\rho^2 = \frac{1}{1-\Omega^2 l^2}$, where, $\Omega$ is the angular velocity of killing vector given by $\Omega = \frac{a_i}{\rho l^2}$.
The general relativistic form of the source-free Maxwell's equations bestow the gauge potential corresponding to the metric, given as
\begin{align}
A_\mu = - \sqrt{\frac{2(n-1)}{n-2}} \frac{q}{r^{n-2}}\left(\rho\,d t - \sum_{i=1}^k a_i \,d \phi_i \right).
\end{align}
The temperature of event horizon by analytic continuation of the metric is used to evaluate the Hawking temperature of the rotating charged quasi-topological black hole, be expressed as \cite{Dehghani2006, Dehghani2011surface, Upadhyay2017quantum}
\begin{align}
\label{reqhtemp}
T_H = \left.\frac{f'(r)}{4 \pi \rho}\right|_{r=r_+} = \frac{nr_+}{4\pi l^2\rho} - \frac{\left(n-2\right)^2 \Phi^2\rho}{8\pi\left(n-1\right)r_+} = 
\frac{nr_+-\left(n-2\right)q^2l^2r_+^{3-2n}}{4\pi l^2\rho},
\end{align}
where electric potential $\Phi$ for static case at infinity with respect to horizon is given by 
\begin{equation}
\label{rep}
\Phi = \sqrt{\frac{2\left(n-1\right)}{\left(n-2\right)}} \frac{q}{\rho r^{n-2}_+}.
\end{equation}
Based on these evaluations, one may obtain the horizon radius in terms of the intensive quantities as follows
\begin{align}
r_+ &= \rho \left[ \frac{2\pi l^2}{n}T_H + 2\left( \frac{\pi^2 l^4}{n^2}T_H^2 + \frac{(n-2)^2 l^2}{8n(n-1)}\Phi^2 \right)^{1/2} \right].
\end{align}
The uncorrected entropy density of charged rotating quasi-topological black hole in the absence of any thermal fluctuation using Gibbs free energy and temperature has been evaluated in the form of
\begin{equation}
\label{reqentropy}
S_0 = \frac{\rho}{4}r^{n-1}_+.
\end{equation}
Thus, the non-perturbatively modified entropy density of charged rotating quasi-topological black hole due to the presence of thermal fluctuation is
\begin{align}
\label{reqentropy}
S &= \frac{\rho}{4}r^{n-1}_+ + \eta e^{-\frac{\rho}{4}r^{n-1}_+}.
\end{align}
\begin{figure}[ht]
     \centering
     \begin{subfigure}[b]{0.45\textwidth}
         \centering
         \includegraphics[width=\textwidth]{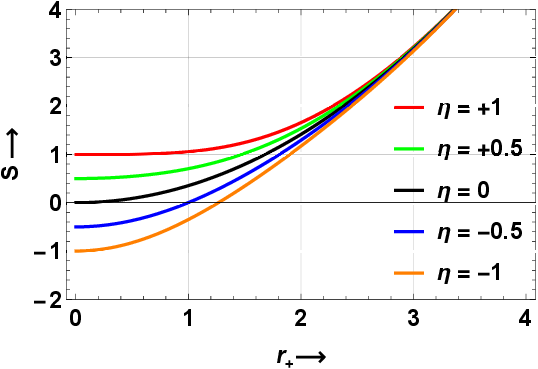}
         \caption{For $n = 3$}
    \label{rentropy3}
     \end{subfigure}
     \hfill
     \begin{subfigure}[b]{0.45\textwidth}
         \centering
         \includegraphics[width=\textwidth]{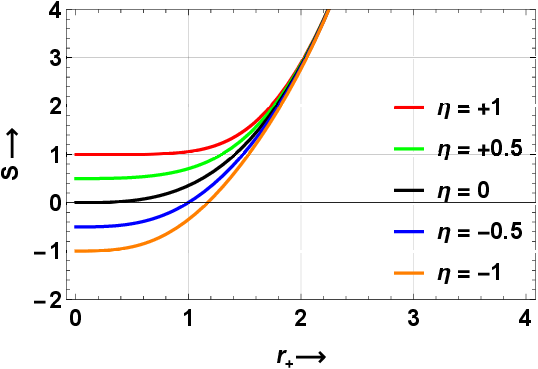}
         \caption{For $n = 4$ }
         \label{rentropy4}
     \end{subfigure} 
           \caption{Variation of entropy per volume $(S)$ with respect to black hole horizon radius $(r_+)$ for  $\rho^2 =2$ at different values of correction parameter $(\eta)$.}
\label{rentropy} 
\end{figure} 
Fig. \ref{rentropy} depicts the variation of uncorrected and non-perturbatively corrected entropy density in four and five space-time dimensions for a comparable study due to the thermal fluctuations.
This monotonically increasing variation shows similar behaviour for uncorrected and corrected entropy with positive correction parameters. In small values of horizon radius, positive correction parameters have positive values and negative correction parameters have negative values. Besides, for the sufficiently large horizon radius of quasi-topological black hole, both positive and negative correction parameters are futile and converge with the equilibrium entropy per volume without any correction parameters.
\subsection{Internal energy}
The exponentially corrected expression for internal energy of this charged rotating quasi-topological black hole is calculated by
\begin{align}
\label{reqinenergy}\nonumber
E & = \int T_H \,d S, \\\nonumber
&=\frac{1}{16 \pi l^2}\left[\left(n-1\right)r^n_+ - \frac{1}{2} \left(n-2\right) \rho^2\phi^2 l^2 r_+^{n-2} + \eta \left\{ n \left(\frac{4}{\rho}\right)^{{\frac{n}{n-1}}} \Gamma\left[\frac{n}{n-1}, \frac{\rho r_+^{n-1}}{4}\right]  \right.  \right. \\& - \left. \left. \frac{\left(n-2 \right)^2}{2\left(n-1 \right)} \rho^2\phi^2 l^2 \left(\frac{4}{\rho}\right)^{{\frac{n-2}{n-1}}} \Gamma\left[\frac{n-2}{n-1}, \frac{\rho r_+^{n-1}}{4}\right] \right\} \right].
\end{align}
\begin{figure}[ht]
     \centering
     \begin{subfigure}[b]{0.45\textwidth}
         \centering
         \includegraphics[width=\textwidth]{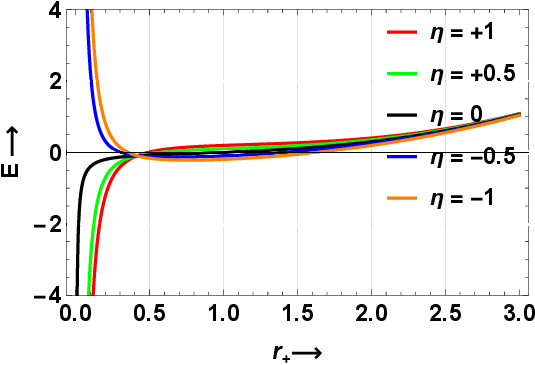}
         \caption{For $n = 3, q = 1, l = 1$}
    \label{renergy311}
     \end{subfigure}
     \hfill
     \begin{subfigure}[b]{0.45\textwidth}
         \centering
         \includegraphics[width=\textwidth]{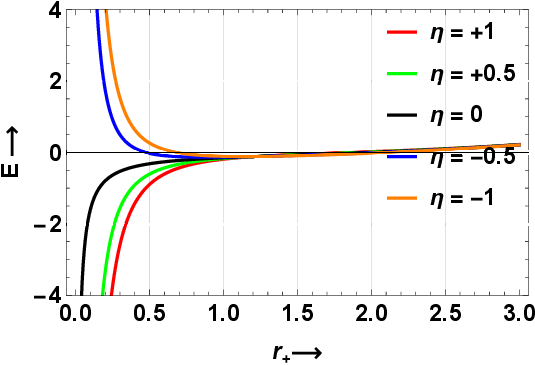}
         \caption{For $n = 3, q = 2, l = 2$ }
         \label{renergy322}
     \end{subfigure}\\
      \begin{subfigure}[b]{0.45\textwidth}
         \centering
         \includegraphics[width=\textwidth]{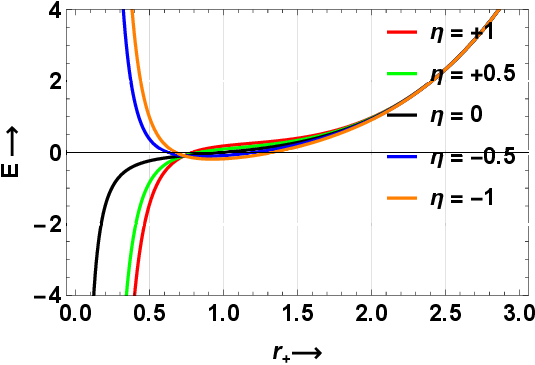}
         \caption{For $n = 4, q = 1, l = 1$}
       \label{renergy411}
     \end{subfigure}
     \hfill
     \begin{subfigure}[b]{0.45\textwidth}
         \centering
         \includegraphics[width=\textwidth]{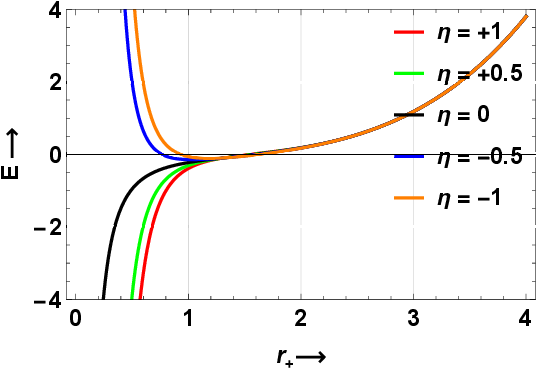}
         \caption{For $n = 4, q = 2, l = 2$}
        \label{renergy422}
     \end{subfigure}
           \caption{Variation of internal energy $(E)$ with respect to black hole horizon radius $(r_+)$ for  $\rho^2$=2  at different values of correction parameter $(\eta)$.}
\label{renergy} 
\end{figure}  
Fig. \ref{renergy} facilitates a comparable analysis of the evolution of exponentially corrected internal energy per volume as well as uncorrected internal energy per volume for the charged quasi-topological black hole.
The variation shows a critical point in lower horizon radius at $r_+=r$. Both in four and five space-time dimensions, positive correction parameters follow the same behavior as followed by uncorrected internal energy in all ranges of horizon radius. Contrariwise, in the lower horizon radius for negative correction parameters, internal energy follows the opposite positively valued asymptote with respect to internal energy for positive correction parameters. 
It is also worth noting that for quasi-topological black hole with a large-valued horizon radius, internal energy in view of the exponential factor behaves more closely to the uncorrected one. Subsequently, these variations converge and exhibit a monotonically increasing trend.
With the increase of dimension $(n)$, charge $(q)$, and AdS radius $(l)$, the converging point of the curve shifted to lower value of horizon radius.
\subsection{Gibbs free energy}
One of the quantum correction on thermodynamical variables is to examine the Gibbs free energy per volume for the quasi-topological black hole in rotating case. 
Following standard definition, this   is given by
\begin{align}
\label{reqgibbs} \nonumber
G(T_H, \Phi, \Omega) &=-\int S \,d T_H, \\ \nonumber
& = - \frac{1}{16\pi l^2} \left[ r_+^n + \frac{(n-2)}{2(n-1)} \rho^2\phi^2 l^2 r_+^{n-2} - \eta\left\{ \frac{n}{n-1} \left(\frac{4}{\rho}\right)^{{\frac{n}{n-1}}} \Gamma\left[\frac{1}{n-1}, \frac{\rho r_+^{n-1}}{4}\right]\right. \right. \\&\left. \left. + \frac{(n-2)^2}{2(n-1)^2} \phi^2 l^2 4^{{\frac{n-2}{n-1}}} \rho^{{\frac{n}{n-1}}} \Gamma\left[\frac{-1}{n-1}, \frac{\rho r_+^{n-1}}{4}\right] \right\} \right], \nonumber\\
&= - \frac{1}{16\pi} \left[ m - \eta\left\{ \frac{n }{(n-1) l^2} \left(\frac{4}{\rho}\right)^{{\frac{n}{n-1}}} \Gamma\left[\frac{1}{n-1}, \frac{\rho r_+^{n-1}}{4}\right]  \right. \right. \\&\left. \left. 
+ \frac{(n-2) q^2 }{(n-1)r_+^{2n-4}} \left(\frac{4}{\rho}\right)^{{\frac{n-2}{n-1}}} \Gamma\left[\frac{-1}{n-1}, \frac{\rho r_+^{n-1}}{4}\right] \right\} \right].
\end{align}
\begin{figure}[ht]
     \centering
     \begin{subfigure}[b]{0.45\textwidth}
         \centering
         \includegraphics[width=\textwidth]{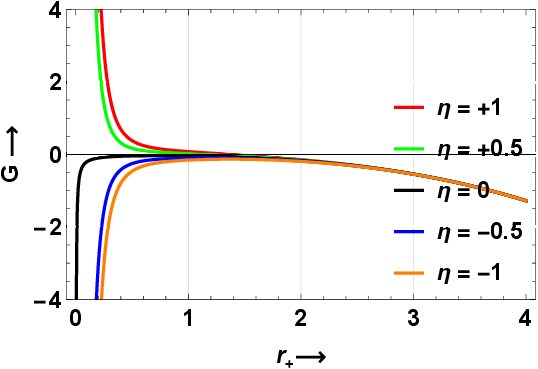}
         \caption{For $n = 3, q = 1, l = 1$}
    \label{rgibbs311}
     \end{subfigure}
     \hfill
     \begin{subfigure}[b]{0.45\textwidth}
         \centering
         \includegraphics[width=\textwidth]{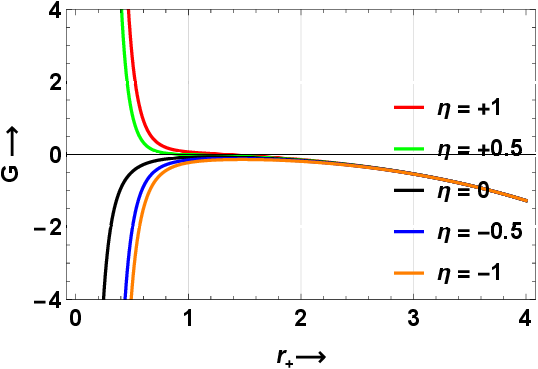}
         \caption{For $n = 3, q = 2, l = 2$ }
         \label{rgibbs322}
     \end{subfigure}\\
      \begin{subfigure}[b]{0.45\textwidth}
         \centering
         \includegraphics[width=\textwidth]{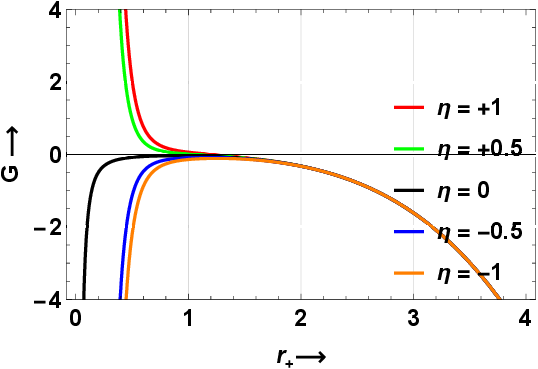}
         \caption{For $n = 4, q = 1, l = 1$}
       \label{rgibbs411}
     \end{subfigure}
     \hfill
     \begin{subfigure}[b]{0.45\textwidth}
         \centering
         \includegraphics[width=\textwidth]{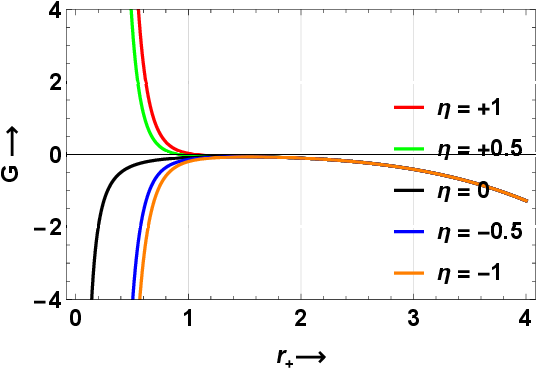}
         \caption{For $n = 4, q = 2, l = 2$}
        \label{rgibbs422}
     \end{subfigure}
           \caption{Variation of Gibbs free energy per volume $(G)$ with respect to black hole horizon radius $(r_+)$ for $\rho^2$=2  at different values of correction parameter $(\eta)$.}
\label{rgibbs} 
\end{figure} 
We plot  Fig. \ref{rgibbs}  for the comparative study of exponentially corrected and uncorrected Gibbs free energy per volume with respect to horizon radius in four and five space-time dimensions.
Gibbs free energy density for positive correction parameters follows the positive asymptotic behavior for small horizon radius. On the contrary, Gibbs free energy for negative and zero valued correction parameters follows the negative asymptotic behavior. After the variation converges, corrected and uncorrected Gibbs free energy density have a negative-valued monotonically increasing function characteristic.
Changes in dimension, charge, and AdS radius do not significantly affect the variations of Gibbs free energy per volume for small horizon radius.
The increased dimension causes a steeper variation of Gibbs free energy with respect to the sufficiently large horizon radius. Conversely, as the charge and AdS radius increase, the variation becomes less steep. 
\subsection{Charge density}
\begin{figure}[ht]
     \centering
     \begin{subfigure}[b]{0.45\textwidth}
         \centering
         \includegraphics[width=\textwidth]{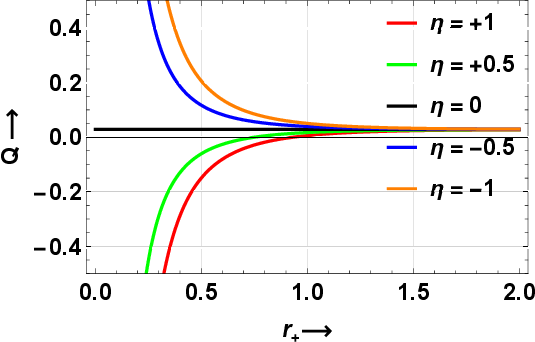}
         \caption{For $n = 3, q = 1, l = 1$}
    \label{rcharge31}
     \end{subfigure}
     \hfill
     \begin{subfigure}[b]{0.45\textwidth}
         \centering
         \includegraphics[width=\textwidth]{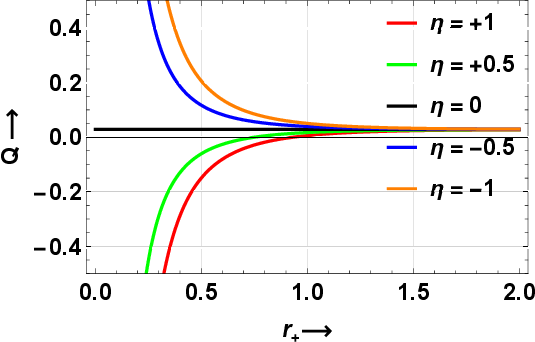}
         \caption{For $n = 3, q = 2, l = 2$ }
         \label{rcharge32}
     \end{subfigure}\\
      \begin{subfigure}[b]{0.45\textwidth}
         \centering
         \includegraphics[width=\textwidth]{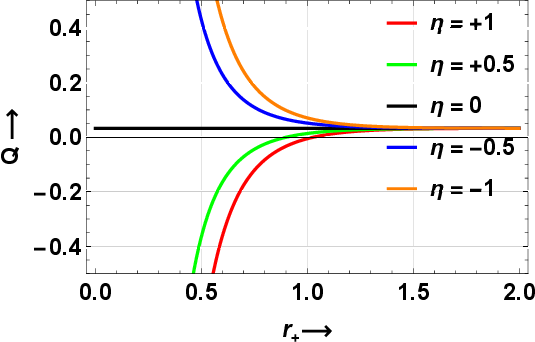}
         \caption{For $n = 4, q = 1, l = 1$}
       \label{rcharge41}
     \end{subfigure}
     \hfill
     \begin{subfigure}[b]{0.45\textwidth}
         \centering
         \includegraphics[width=\textwidth]{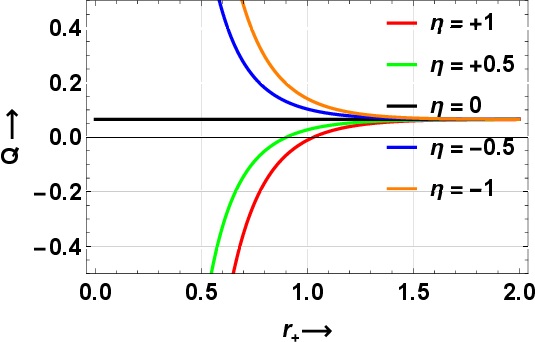}
         \caption{For $n = 4, q = 2, l = 2$}
        \label{rcharge42}
     \end{subfigure}
           \caption{Variation of Charge $(Q)$ with respect to black hole horizon radius $(r_+)$ for   $\rho^2$=2  at different values of correction parameter $(\eta)$.}
\label{rCharge} 
\end{figure}

In this subsection, corrected charged density of charged rotating quasi-topological black holeunder the influence of  exponential correction due to the thermal fluctuations is given by
\begin{align}
\label{reqcharge}
Q &= \frac{1}{16 \pi} \sqrt{\frac{2(n-2)}{(n-1)}}q \left\{ \rho - \eta  \frac{(n-2)4^{{\frac{n-2}{n-1}}} \rho^{{\frac{1}{n-1}}}}{(n-1)r_+^{n-2}} \Gamma\left[\frac{-1}{n-1}, \frac{\rho r_+^{n-1}}{4}\right] \right\}.
\end{align}
This expression can be reduced in form of total equilibrium charge density on taking the limit $\eta \rightarrow 0$ as in Ref. \cite{Dehghani2011surface}. 
The most remarkable point of this expression is that total charge for this charged rotating quasi-topological black hole is non-existent for less than three space dimensions. 
The graphical variations of corrected and uncorrected charge density in four and five space-time dimensions are shown in Fig. \ref{rCharge}.
The positive and negative correction parameters impart negative and positive opposite asymptotic behavior for small horizon radius. For large horizon radius, effect of correction parameters gets worn off.

\subsection{Angular momentum}
Since the charged rotating quasi-topological black hole incorporates rotation, it is necessary to include another extensive parameter, namely the angular momentum per volume, which may be calculated using a non-perturbative corrected parameter $(\eta)$ as follows
\begin{align}
\label{reqam}\nonumber
J_i &= -\left(\frac{\partial G}{\partial \Omega_i}\right)_{T, \Phi},
\\&=
\frac{a_i \rho}{16 \pi} \left(nm - \frac{2q^2}{r_+^{n-2}}\right) \left( 1+ \frac{4 \eta}{\rho r_+^{n-1}} e^{-\frac{\rho r_+^{n-1}}{4}} \right).
\end{align}
In the absence of thermal fluctuation, charge-independent angular momentum can be reduced to equilibrium form, as evaluated in Ref. \cite{Dehghani2011surface}.
\subsection{Mass density}
The corrected mass density as a function of extensive parameters such as entropy, angular momentum, and charge, of this charged rotating quasi-topological black hole is given by
\begin{align}
\label{reqmass1} \nonumber
M &= G + TS + \Phi Q + \sum_{i=1}^k \Omega_i J_i \\ \nonumber
& = \frac{1}{16 \pi l^2} \left[ (n-1)r_+^n +\frac{q^2 l^2 (3-n)}{r_+^{n-2}} + l^2(\rho^2-1) \left( \frac{n r_+^n} {l^2}+ \frac{(n-2)q^2}{r_+^{n-2}} \right)
\right. \\\nonumber& \left.
+ \eta \left\{ \frac{n}{n-1} \left(\frac{4}{\rho}\right)^{{\frac{n}{n-1}}} \Gamma\left[\frac{1}{n-1}, \frac{\rho r_+^{n-1}}{4}\right] 
-\frac{(n-2)q^2 l^2}{(n-1)r_+^{2n-4}}\left(
\frac{4}{\rho}\right)^{{\frac{n-2}{n-1}}} \Gamma\left[\frac{-1}{n-1}, \frac{\rho r_+^{n-1}}{4}\right] \right. \right. \\& \left. \left.
+ \frac{4l^2}{\rho} e^{-\frac{\rho r_+^{n-1}}{4}} \left\{ \left( \frac{nr_+}{l^2} - \frac{(n-2)q^2}{r_+^{2n-3}} \right) + (\rho^2-1) \left( \frac{n r_+^n}{l^2} + \frac{(n-2)q^2}{r_+^{2n-3}} \right)  \right\}
\right\} \right].
\end{align}
\begin{figure}[ht]
     \centering
     \begin{subfigure}[b]{0.45\textwidth}
         \centering
         \includegraphics[width=\textwidth]{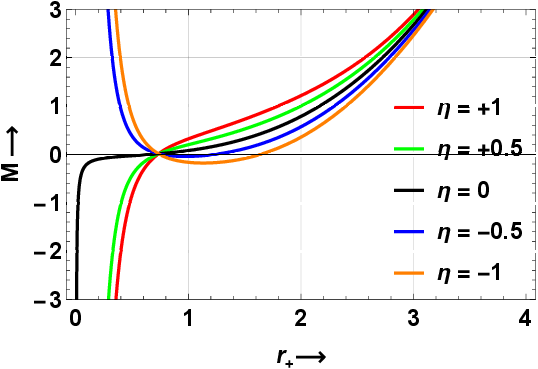}
         \caption{For $n = 3, q = 1, l = 1$}
    \label{rmass311}
     \end{subfigure}
     \hfill
     \begin{subfigure}[b]{0.45\textwidth}
         \centering
         \includegraphics[width=\textwidth]{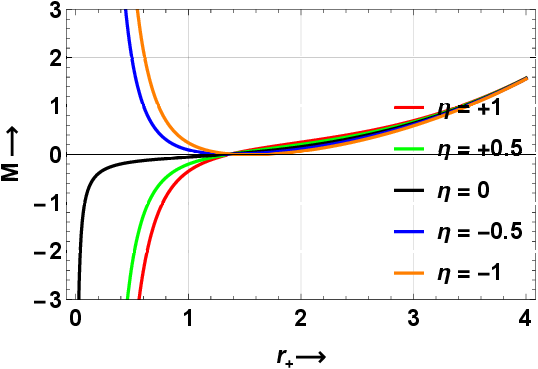}
         \caption{For $n = 3, q = 2, l = 2$ }
         \label{rmass322}
     \end{subfigure}\\
      \begin{subfigure}[b]{0.45\textwidth}
         \centering
         \includegraphics[width=\textwidth]{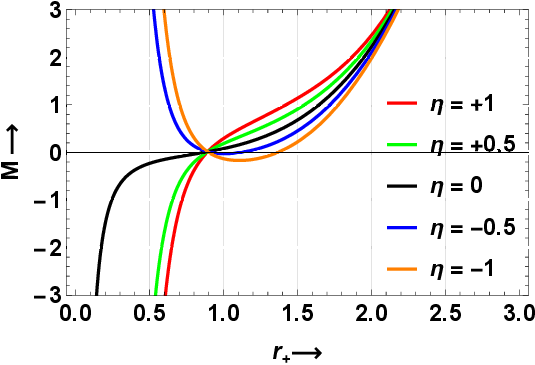}
         \caption{For $n = 4, q = 1, l = 1$}
       \label{rmass411}
     \end{subfigure}
     \hfill
     \begin{subfigure}[b]{0.45\textwidth}
         \centering
         \includegraphics[width=\textwidth]{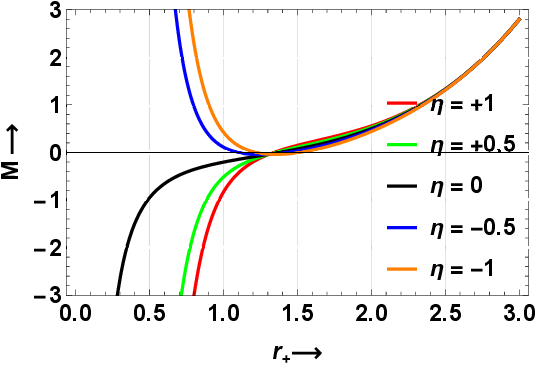}
         \caption{For $n = 4, q = 2, l = 2$}
        \label{rmass422}
     \end{subfigure}
           \caption{Variation of mass per volume $(M)$ with respect to black hole horizon radius $(r_+)$ for  $\rho^2=2$ at different values of correction parameter $(\eta)$.}
\label{rmass} 
\end{figure} 
 The corrected mass in term of mass of rotating quasi-topological black hole is
\begin{align}
\label{reqmass2} \nonumber
M & = \frac{1}{16 \pi l^2} \left[ l^2 \left\{(n-1)m  +\frac{q^2 (4-2n)}{r_+^{n-2}} \right\} + l^2(\rho^2-1) \left( nm- \frac{2q^2}{r_+^{n-2}} \right)
\right.\\ \nonumber
& \left.
+ \eta \left\{\frac{n}{n-1} \left(\frac{4}{\rho}\right)^{{\frac{n}{n-1}}} \Gamma\left[\frac{1}{n-1}, \frac{\rho r_+^{n-1}}{4}\right] 
-\frac{(n-2)q^2 l^2}{(n-1)r_+^{2n-4}}\left(
\frac{4}{\rho}\right)^{{\frac{n-2}{n-1}}} \Gamma\left[\frac{-1}{n-1}, \frac{\rho r_+^{n-1}}{4}\right] \right. \right. \\& \left. \left.
+ \frac{4l^2}{\rho} e^{-\frac{\rho r_+^{n-1}}{4}} \left\{ \left( \frac{nr_+}{l^2} - \frac{(n-2)q^2}{r_+^{2n-3}} \right) + (\rho^2-1) \left( \frac{n r_+^n}{l^2} + \frac{(n-2)q^2}{r_+^{2n-3}} \right)  \right\}
\right\} \right].
\end{align}
The total mass density expression can be reduced for equilibrium mass without any thermal fluctuations by taking the limit $\eta\rightarrow 0$ as in Ref. \cite{Dehghani2011surface}. 
The variation in three and four space dimensions for a comparable study of corrected and uncorrected mass density with respect to horizon radius is shown in Fig. \ref{rmass}. From these variations, it can be inferred that we are experiencing a critical point with a small horizon radius at $r_+=r$. Below the critical point, in lower range of horizon radius, uncorrected and corrected mass per volume with positive correction parameter shows the negative asymptote of similar nature. 
Whereas, negative correction parameters impart the positive asymptote. 
For the large value of horizon radius, uncorrected and both positively and negatively corrected parameters converge and follow the monotonically increasing function with the positive value of mass per volume.
\subsection{Specific heat and Stability}
The stability of black hole thermodynamic system due to the thermal fluctuations is based on the analysis of sign of specific heat. The root value defines the phase transition from negatively valued unstable state to the positively valued stable state of the black hole.
The specific heat per unit volume for charged rotating quasi-topological black hole with a fixed chemical potential $(\Phi)$ is as follows
\begin{figure}[ht]
     \centering
     \begin{subfigure}[b]{0.45\textwidth}
         \centering
         \includegraphics[width=\textwidth]{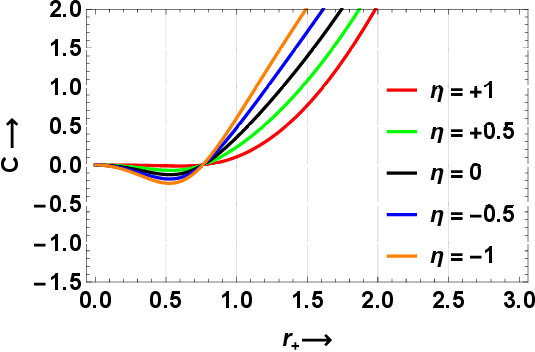}
         \caption{For $n = 3, q = 1, l = 1$}
    \label{rsh311}
     \end{subfigure}
     \hfill
     \begin{subfigure}[b]{0.45\textwidth}
         \centering
         \includegraphics[width=\textwidth]{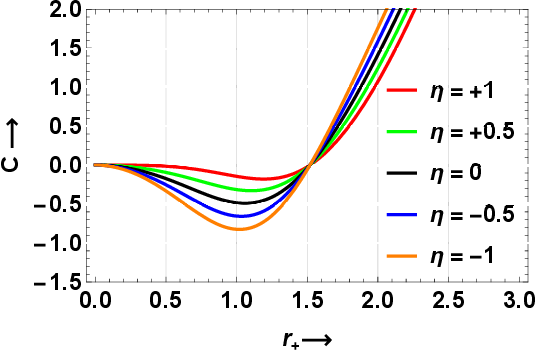}
         \caption{For $n = 3, q = 2, l = 2$ }
         \label{rsh322}
     \end{subfigure}\\
      \begin{subfigure}[b]{0.45\textwidth}
         \centering
         \includegraphics[width=\textwidth]{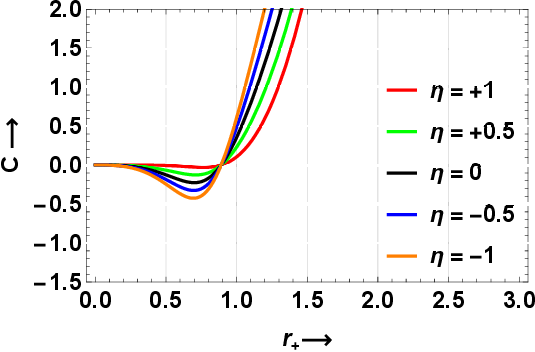}
         \caption{For $n = 4, q = 1, l = 1$}
       \label{rsh411}
     \end{subfigure}
     \hfill
     \begin{subfigure}[b]{0.45\textwidth}
         \centering
         \includegraphics[width=\textwidth]{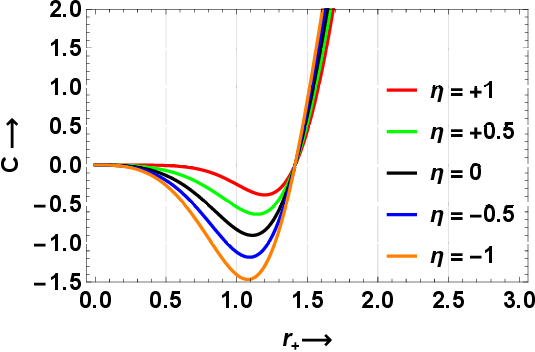}
         \caption{For $n = 4, q = 2, l = 2$}
        \label{rsh422}
     \end{subfigure}
           \caption{Variation of Specific heat per volume $(C)$ with respect to black hole horizon radius $(r_+)$ for   $\rho^2$=2   at different values of correction parameter $(\eta)$.}
\label{rsh} 
\end{figure} 
\begin{align}\nonumber
C_\Phi & = \frac{2\pi \rho^2 l^2 r_+^n (n-1)^2 T}{2n(n-1)r_+^2 +(n-2)^2 \rho^2 \phi^2 l^2}\left[1-\eta e^{-\frac{\rho r_+^{n-1}}{4}}\right],\\\label{reqsh}
& = \frac{\rho\left(n-1\right)r_+^{n-1} \left(\frac{2n}{l^2}- \frac{(n-2)^2\phi^2\rho^2}{(n-1)r_+^2}\right)}{4 \left(\frac{2n}{l^2}+\frac{(n-2)^2\phi^2 \rho^2}{(n-1)r_+^2}\right)}\left[1-\eta e^{-\frac{\rho r_+^{n-1}}{4}}\right].
\end{align}
On solving the Eq.(\ref{reqsh}) to determine bound points with respect to horizon radius, for the case of stability, one may have
\begin{align}
\label{reqstability}
r_{c} = \left[-\frac{n-2}{n}q^2 l^2 \right]^{1/2n-2}.
\end{align}
Fig. \ref{rsh} shows the comparable study of corrected and uncorrected specific heat per volume with respect to horizon radius. The variations for different correction parameters possess a critical point at $r_+=r$. Below the range of this critical point, both corrected and uncorrected specific heat have negative valued specific heat and above the critical point value, all have positive value of specific heat per volume. The variations are convex functions and follow the monotonically increasing nature. The critical point provides the phase transition from unstable state to stable state of the charged rotating quasi-topological black hole. For the large value of horizon radius, the correction parameter becomes insignificant. 
\section{Conclusions}
\label{last}
This work detailed the gravitational theory under the influence of Gauss-Bonnet term with curvature cubed interactions for $(n+1)$ space-time dimensions in the presence of an electromagnetic field as a quasi-topological gravity. The charged and charged rotating quasi-topological black hole gravity, which corresponds to the exact black hole solutions, have been studied to analyse the effect on different thermodynamical parameters due to the thermal fluctuations.
The term $(\rho)$ gets implanted as a function of angular velocity and killing vector in the thermodynamical parameters associated with rotating quasi-topological black hole.
For the sufficient decrease in the size of black hole due to the Hawking radiation, non-perturbative analysis dominates in the form of an exponential correction.
The exponential corrections are found to be negligible at large horizon radius but impose a high impact in the thermodynamics of this system at small horizon radius.
 
The effects of positive and negative correction parameters in non-perturbative analysis of entropy density have been examined in accordance with the uncorrected entropy densities and plotted with respect to the horizon radius. The  positive and negative correction parameters exhibit positive and negative entropy density, respectively, in the lower horizon radius. 
These variations converge at large horizon radius and follow the monotonically increasing behaviour without any substantial effect of the correction parameter. 
The negative value of corrected entropy per volume does not correspond to the physical condition. The internal energy of this charged quasi-topological black hole system is computed from the first-law of thermodynamics.
Furthermore, Gibbs free energy per volume provides the opposite asymptote of similar nature as of the correction parameters' signature in lower region. The negative correction parameters follow the uncorrected Gibbs free energy variation.
The corrected charge adheres to the similar nature as the correction parameter and remains independent with respect to horizon radius for uncorrected charge.
The exponentially corrected internal energy and mass density with positive and negative correction parameters have been comparably examined with the uncorrected internal energy. 
The product of angular velocity and angular momentum has been introduced as an additional term for the calculation of mass per volume in charged rotating quasi-topological black hole system.
This variation provides the opposite asymptotes and of contrary behaviour for positive and negative correction parameters in small range of horizon radius.
Motivated by these analysis, we have employed the non-perturbative corrections with positive and negative parameters with the specific heat density.
Below the critical point, both corrected and uncorrected specific heat per volume is in unstable state. From there, it follows the first order phase transition from negative valued specific heat to positive valued specific heat in the stable state.
On solving the denominator of specific heat, singularity which defines the bound points with respect to horizon radius, to discuss the stability of the system, provides the exactly similar result for charged rotating quasi-topological black hole as determined in the case of non-rotating charged quasi-topological black hole.
\section*{Acknowledgement} 
This research was funded by the Science Committee of the Ministry of Science and Higher Education of the Republic of Kazakhstan (Grant No. AP22682760). D.V.S. would like to thank DST-SERB for the project no. EEQ/2022/00824.
 
\bibliographystyle{ieeetr}
\bibliography{library}

\begin{thebibliography}{10}

\bibitem{Maldacena1999}
J.~Maldacena, ``{The large-N limit of superconformal field theories and
  supergravity},'' {\em Int. J. Theor. Phys}, vol.~38, no.~4, pp.~1113--1133,
  1999.

\bibitem{Witten1998}
E.~Witten, ``{Anti de sitter space and holography},'' {\em ATMP}, vol.~2,
  no.~2, pp.~253--290, 1998.

\bibitem{Gubser1998}
S.~S. Gubser, I.~R. Klebanov, and A.~M. Polyakov, ``{Gauge theory correlators
  from non-critical string theory},'' {\em Phys. Lett. B}, vol.~428, no.~1-2,
  pp.~105--114, 1998.

\bibitem{Buchel2005}
A.~Buchel, J.~T. Liu, and A.~O. Starinets, ``{Coupling constant dependence of
  the shear viscosity in N = 4 supersymmetric Yang-Mills theory},'' {\em Nucl.
  Phys. B}, vol.~707, no.~1-2, pp.~56--68, 2005.

\bibitem{Myers2010holographic}
R.~C. Myers, M.~F. Paulos, and A.~Sinha, ``{Holographic studies of
  quasi-topological gravity},'' {\em Journal of High Energy Physics},
  vol.~2010, no.~8, pp.~1--44, 2010.

\bibitem{Hofman2008}
D.~M. Hofman and J.~Maldacena, ``{Conformal collider physics: Energy and charge
  correlations},'' {\em J. High Energy Phys.}, vol.~2008, no.~5, pp.~1--73,
  2008.

\bibitem{Hofman2009}
D.~M. Hofman, ``{Higher derivative gravity, causality and positivity of energy
  in a UV complete QFT},'' {\em Nucl. Phys. B}, vol.~823, no.~1-2,
  pp.~174--194, 2009.

\bibitem{Myers2010black}
R.~C. Myers and B.~Robinson, ``{Black holes in quasi-topological gravity},''
  {\em Journal of High Energy Physics}, vol.~2010, no.~8, pp.~1--34, 2010.

\bibitem{Oliva2010}
J.~Oliva and S.~Ray, ``{A new cubic theory of gravity in five dimensions :
  Black hole, Birkhoff's theorem and C-function},'' {\em Class. Quant. Grav.},
  vol.~27, pp.~1--21, 2010.

\bibitem{Brenna2012}
W.~G. Brenna and R.~B. Mann, ``{Quasitopological Reissner-Nordstr{\"{o}}m black
  holes},'' {\em Phys. Rev. D}, vol.~86, no.~6, pp.~1--12, 2012.

\bibitem{Dehghani2011surface}
M.~H. Dehghani and M.~H. Vahidinia, ``{Surface terms of quasitopological
  gravity and thermodynamics of charged rotating black branes},'' {\em Phys.
  Rev. D}, vol.~84, no.~8, pp.~1--13, 2011.

\bibitem{Bardeen1973}
J.~M. Bardenn, B.~Carter, and S.~W. Hawking, ``{The four laws of black hole
  mechanics},'' {\em Commun. math. Phys.}, vol.~31, pp.~161--170, 1973.

\bibitem{Bekenstein1973}
J.~D. Bekenstein, ``{Black holes and entropy},'' {\em Phys. Rev. D}, vol.~7,
  no.~8, pp.~2333--2346, 1973.

\bibitem{Hawking1976}
S.~W. Hawking, ``{Black holes and thermodynamics},'' {\em Phys. Rev. D},
  vol.~13, no.~2, pp.~191--197, 1976.

\bibitem{Mathur2012}
S.~D. Mathur, ``{Black holes and holography},'' {\em J. Phys.: Conf. Ser},
  vol.~405, no.~1, pp.~1--9, 2012.

\bibitem{Pourhassan2021a}
B.~Pourhassan, M.~Dehghani, M.~Faizal, and S.~Dey, ``{Non-perturbative quantum
  corrections to a Born-Infeld black hole and its information geometry},'' {\em
  Class. Quant. Grav.}, vol.~38, no.~10, pp.~1--17, 2021.

\bibitem{Pourhassan2021b}
B.~Pourhassan and M.~Faizal, ``{Quantum corrections to the thermodynamics of
  black branes},'' {\em JHEP}, vol.~2021, no.~10, 2021.

\bibitem{Gulin2018}
L.~Gulin and I.~Smoli{\'{c}}, ``{Generalizations of the Smarr formula for black
  holes with nonlinear electromagnetic fields},'' {\em Class. Quantum Grav.},
  vol.~35, no.~2, pp.~1--18, 2018.

\bibitem{Page2005}
D.~N. Page, ``{Hawking radiation and black hole thermodynamics},'' {\em New J.
  Phys.}, vol.~7, pp.~1--46, 2005.

\bibitem{Ashtekar2020}
A.~Ashtekar, ``{Black hole evaporation: A perspective from loop quantum
  gravity},'' {\em Universe}, vol.~6, no.~2, pp.~1--25, 2020.

\bibitem{Upadhyay2018}
S.~Upadhyay, S.~H. Hendi, S.~Panahiyan, and B.~E. Panah, ``{Thermal
  fluctuations of charged black holes in gravity's rainbow},'' {\em Prog.
  Theor. Exp. Phys.}, vol.~2018, no.~9, pp.~1--20, 2018.

\bibitem{Dehghani2019}
M.~Dehghani, ``{Thermal fluctuations of AdS black holes in three-dimensional
  rainbow gravity},'' {\em Phys. Lett. B}, vol.~793, pp.~234--239, 2019.

\bibitem{PourhassanUpadhyay2021}
B.~Pourhassan and S.~Upadhyay, ``{Thermal fluctuations of charged black hole
  solution in Rastall theory},'' {\em Eur. Phys. J. Plus}, vol.~136, no.~311,
  pp.~1--23, 2021.

\bibitem{Pourhassan2021}
B.~Pourhassan, ``{Exponential corrected thermodynamics of black holes},'' {\em
  J. Stat. Mech.: Theory Exp.}, vol.~2021, no.~7, pp.~1--29, 2021.

\bibitem{Upadhyay2017quantum}
S.~Upadhyay, ``{Quantum corrections to thermodynamics of quasitopological black
  holes},'' {\em Phys. Lett. B}, vol.~775, no.~2017, pp.~130--139, 2017.

\bibitem{Chatterjee2020}
A.~Chatterjee and A.~Ghosh, ``{Exponential Corrections to Black Hole
  Entropy},'' {\em Phys. Rev. Lett.}, vol.~125, no.~4, pp.~1--6, 2020.

\bibitem{Soroushfar2023}
S.~Soroushfar, H.~Farahani, and S.~Upadhyay, ``{Non-perturbative correction to
  thermodynamics of conformally dressed 3D black hole},'' {\em Physics of Dark
  Universe}, vol.~42, pp.~1--12, 2023.

\bibitem{non2024pourhassan}
B.~Pourhassan, H.~Farahani, F.~Kazemian, and i.~Sakalli, ``{Non-perturbative
  correction on the black hole geometry},'' {\em Physics of the Dark Universe},
  vol.~44, pp.~1--9, 2024.

\bibitem{Xu2020}
H.~Xu and Y.~C. Ong, ``{Black hole evaporation in Horava--Lifshitz gravity},''
  {\em Eur. Phys. J. C}, vol.~80, no.~679, pp.~1--10, 2020.

\bibitem{Emparan2023}
R.~Emparan, E.~Barakovic, R.~Dekhil, and F.~Rescic, ``{Black holes in the
  classical and quantum world},'' {\em PoS Proc. Sci.}, pp.~1--51, 2023.

\bibitem{Pourhassan2022}
B.~Pourhassan and İzzet Sakallı, ``Non-perturbative correction to the
  hořava–lifshitz black hole thermodynamics,'' {\em Chinese Journal of
  Physics}, vol.~79, pp.~322--338, 2022.

\bibitem{Rama1999}
S.~K. Rama, ``{Holographic Principle in the Closed Universe : A Resolution with
  Negative Pressure Matter},'' {\em Phys. Lett. B}, vol.~457, no.~4,
  pp.~268--274, 1999.

\bibitem{Bak2000}
D.~Bak and S.~J. Rey, ``{Holographic principle and string cosmology},'' {\em
  Class. Quantum Grav.}, vol.~17, no.~1, pp.~L1--L7, 2000.

\bibitem{Kaul2000}
R.~K. Kaul and P.~Majumdar, ``{Logarithmic correction to the bekenstein-hawking
  entropy},'' {\em Phys. Rev. Lett.}, vol.~84, no.~23, pp.~5255--5257, 2000.

\bibitem{Dabholkar2011}
A.~Dabholkar, J.~Gomes, and S.~Murthy, ``{Quantum black holes, localization,
  and the topological string},'' {\em JHEP}, vol.~2011, no.~6, pp.~1--51, 2011.

\bibitem{Upadhyay2017}
S.~Upadhyay, ``{Thermodynamics and galactic clustering with a modified
  gravitational potential},'' {\em Phys. Rev. D}, vol.~95, no.~4, pp.~1--10,
  2017.

\bibitem{Iorio2020}
A.~Iorio, G.~Lambiase, P.~Pais, and F.~Scardigli, ``{Generalized uncertainty
  principle in three-dimensional gravity and the BTZ black hole},'' {\em Phys.
  Rev. D}, vol.~101, no.~10, pp.~1--13, 2020.

\bibitem{Pourhassan2018}
B.~Pourhassan, K.~Kokabi, and Z.~Sabery, ``{Higher order corrected
  thermodynamics and statistics of Kerr–Newman–G{\"{o}}del black hole},''
  {\em Annals of Physics}, vol.~399, pp.~181--192, 2018.

\bibitem{Soroushfar2024}
S.~Soroushfar, B.~Pourhassan, and I.~Sakalli, ``{Exploring Non-perturbative
  Corrections in Thermodynamics of Static Dirty Black Holes},'' {\em Physics of
  Dark Universe}, vol.~44, pp.~1--16, 2024.

\bibitem{Bueno2020generalised}
P.~Bueno, P.~A. Cano, and R.~A. Hennigar, ``{(Generalized) quasi-topological
  gravities at all orders},'' {\em Class. Quantum Grav.}, vol.~37, no.~1,
  pp.~1--27, 2020.

\bibitem{Lovelock1971the}
D.~Lovelock, ``{The Einstein tensor and its generalizations},'' {\em Journal of
  Mathematical Physics}, vol.~12, no.~3, pp.~498--501, 1971.

\bibitem{Deruelle1990lovelock}
N.~Deruelle and L.~Fari{\~{n}}a-Busto, ``{Lovelock gravitational field
  equations in cosmology},'' {\em Physical Review D}, vol.~41, no.~12,
  pp.~3696--3708, 1990.

\bibitem{Bueno2024regular}
P.~Bueno, P.~A. Cano, and R.~A. Hennigar, ``{Regular Black Holes From Pure
  Gravity},'' 2024.

\bibitem{Filippo2024inner}
F.~{Di Filippo}, I.~Kol{\'{a}}ř, and D.~Kubiznak, ``{Inner-extremal regular
  black holes from pure gravity},'' pp.~1--12, 2024.

\bibitem{Cvetic1999}
M.~Cveti{\v{c}} and S.~S. Gubser, ``{Phases of R-charged black holes, spinning
  branes and strongly coupled gauge theories},'' {\em J. High Energy Phys.},
  vol.~3, no.~4, pp.~1--30, 1999.

\bibitem{Caldarelli2000}
M.~M. Caldarelli, G.~Cognola, and D.~Klemm, ``{Thermodynamics of
  Kerr-Newman-AdS black holes and conformal field theories},'' {\em Class.
  Quantum Grav.}, vol.~17, no.~2, pp.~399--420, 2000.

\bibitem{Dehghani2006}
M.~H. Dehghani and R.~B. Mann, ``{Thermodynamics of rotating charged black
  branes in third order lovelock gravity and the counterterm method},'' {\em
  Phys. Rev. D}, vol.~73, no.~10, pp.~1--19, 2006.

\bibitem{Awad2019}
A.~M. Awad, G.~G. Nashed, and W.~{El Hanafy}, ``{Rotating charged AdS solutions
  in quadratic f(T) gravity},'' {\em Eur. Phys. J. C}, vol.~79, no.~8,
  pp.~1--11, 2019.

\end{thebibliography}

%
\end{document}